\documentclass[%
 reprint,
 amsmath,amssymb,
 aps,
]{revtex4-2}

\usepackage{graphicx}
\usepackage{dcolumn}
\usepackage{bm}

\renewcommand{\selectlanguage}[1]{}

\begin{document}

\preprint{APS/123-QED}

\title{The Effect of Value-Focused Discussions on Scientists' Ethical Decision Making}

\author{Tyler Garcia}
\affiliation{Physics Department, Kansas State University}

\author{Bill Bridges}
\affiliation{Physics Department, Kansas State University}
 
\author{Caitlin Solis}
\affiliation{Physics Department, Kansas State University}


\author{Caleb Linville}
\affiliation{Philosophy Department, Kansas State University}

\author{Wyatt Jones}
\affiliation{Physics Department, Kansas State University}

\author{Scott Tanona}
\affiliation{Philosophy Department, Kansas State University}

\author{Jonathan Herington}
\affiliation{Department of Health Humanities and Bioethics, University of Rochester}
\altaffiliation[Also at ]{Philosophy Department, Kansas State University}

\author{James T. Laverty}
\affiliation{Physics Department, Kansas State University}


\date{\today}

\begin{abstract}
Many scientists view science as value-free, despite the fact that both epistemic and non-epistemic values structure scientific inquiry. Current ethics training usually focuses on transmitting knowledge about high-level ethical concepts or rules and is widely regarded as ineffective. We argue that ethics training will be more effective at improving ethical decision making if it focuses on connecting values to science. We pull from philosophy and psychology to define ethical decision making using the Four Component Model. This model states that in order to make an ethical decision someone must consider four components: moral sensitivity, moral reasoning, moral motivation, and moral implementation. We formed a moderated fellowship of fourteen science faculty from different disciplines who met for ten sessions over the course of a year, where they discussed the values embedded in different scientific norms. We then conducted interviews before and after the year-long fellowship that involved guided reflection of scenarios where there was some kind of ethical misconduct where the scientific practice required value judgements (e.g using unpublished data in their own work). We looked at how the fellowship affected the scientists' ability to recognize ethical dimensions regarding the scenarios. We found that this fellowship improved moral sensitivity, but their moral reasoning does not improve. We outlined our approach on how to look at scientists' ethical decision making and made recommendations on how to improve our approach. This work can inform future ethical training to align better with what scientists value and introduce useful concepts from philosophy and psychology to education research in physics.
\end{abstract}

\maketitle

\renewcommand{\arraystretch}{1.2}


\section{\label{sec:level1}Introduction}

In 2019 the American Physical Society (APS) updated their Guidelines on Ethics to further clarify physicist's duties to avoid data fabrication, falsification, plagiarism, and abuse  \cite{Guidlines}. These updates were in response to the National Academies' Fostering Integrity in Research report, which identified a need to improve research integrity training due to the growing importance of information technology in research and the number of studies being retracted because of irreproducible data \cite{fostering}. While these guidelines are an important step forward, we suggest that to implement these guidelines appropriately, physicists must cultivate their ethical decision-making ~\cite{reiss_teaching_1999}.

Currently, physicists are primarily exposed to ethics through Responsible Conduct of Research (RCR) training. This kind of training, often offered in self-paced online modules, tends to focus on topics like authorship, conflicts of interest, research misconduct, and the protection of human subjects \cite{CITI}. Studies have shown that this kind of online training is ineffective at improving scientists' ethical decision-making \cite{mumford_read_nodate, powell_effectiveness_2007, antes_meta-analysis_2009}. Specifically in physics, Kirby and Houle found in 2004 that around 39\% of junior members of APS who responded to the survey had observed or had personal knowledge of at least one ethical violation, such as data falsification, plagiarism, or authorship showing that ethical training is needed, in physics\cite{kirby_ethics_2004}. This survey was re-administered in 2020 to junior level physicists and grad students. The survey showed mixed improvement: physicists were undergoing more ethics training (whether at the institution level or with their supervisors), but the number of misconduct violations did not significantly change \cite{houle_ethics_2023}. Those that did notice violations reported more violations than in 2003 \cite{houle_ethics_2023}. The 2020 study also found that harassment or abuse (e.g. experiencing inappropriate remarks, treated differently, etc.) of physicists is a common experience, concentrated among women in physics \cite{houle_ethics_2023}. In sum, there is more to be done to improve physicists' ethical conduct.

In our study of scientists' ethical decision making, we focus on the fact that scientists routinely make value judgements in science. In this paper, we define ``values in science'' to refer to aims, goals, or principles that direct or influence scientific  work and ``value judgements'' to refer to evaluations of priorities between values (e.g., deciding tradeoffs) or of how a value is to be implemented in a decision or action (see Theory section for more detail). The idea that scientists make value judgements in science seems to contrast with the commonly held belief that that social and ethical values ought not affect the agenda, methodology or dissemination of research \cite{reiss_teaching_1999}. However, more recently philosophers of science have argued persuasively that science can never be value free, and pretending otherwise distorts scientist's sense of their own ethical responsibilities \cite{brown_values_2013,douglas_inductive_2000}. Ethics, after this recognition of values in science, requires attention to those values and reasoning well about them.

For instance, Douglas has argued that scientists need to attend to non-epistemic values (e.g. consequences on people's lives) as well as epistemic values (e.g. gaining knowledge) to address the ``inductive risk" of making an error in either accepting or rejecting a claim \cite{douglas_inductive_2000}. Scientists often  report the significance of statistical findings by using a `p-value' to indicate the evidential strength of  their data. The choice of a statistical significance level for a p-value (i.e. p $<$ 0.01, p $<$ 0.05, etc.) implies a tradeoff between the risk of a false positive result and the risk of a false negative result. Moreover, choices in experimental design, and sometimes during experiments themselves, influence the chances of positive or negative results. These tradeoffs are not merely epistemic, as false positives and false negatives can differentially affect peoples lives, directly or cumulatively, both within the scientific community and in the public more broadly (e.g., consider the replication crisis, or the consequences of requiring definitive evidence of a health risk before action). This highlights both that scientists cannot avoid making value judgements and that the scope of ethics in science involves a broader range of topics than is traditionally covered in RCR training. Building on this work, we argue that we can improve physicist's ethical decision making by focusing on the values invoked in actual scientific practice.

In this paper, we define ``ethics" to refer to the domain of inquiry concerning \textit{ultimately doing the right thing} \cite{stokhof_ethics_2018}, and ``ethical decision-making" as a process  (individual or collective) aimed at doing the right thing. While some invoke a distinction between the adjectives ``ethical" and ``moral", in this paper we align with the convention in the philosophical and psychological literature on ethical decision-making that uses these words interchangeably.

The Four Component Model of ethical decision-making  \cite{rest_moral_1994}  identifies four necessary components to making an ethical decision: moral sensitivity (identifying morally important ideas), moral reasoning (identifying morally right action), moral motivation (intending to do the morally right action), and moral implementation (acting on morally right action) \cite{narvaez_four_1995}. (Note that while the original name for the second component is ``moral judgement", in this paper we call the second component ``moral reasoning", consistent with some of the literature on this topic) \cite{you_independence_2013}. In this paper we focused on the moral sensitivity and moral reasoning components in ethical decision making. These components and their definitions will be further explained in the Theory section.

This paper reports on an intervention aimed at learning how to use the presence of value judgements in science to help improve scientists' ethical decision making. This intervention took the form of a two-semester fellowship in which  scientists from different fields met together to talk about scientific topics rooted in value judgements. To see if their ethical decision making changed over the course of the fellowship, we conducted (one-on-one) interviews with the faculty before and after the fellowship to talk about their own experience of research ethics and their responses to vignettes describing situations with multiple ethical concerns identified by RCR training. 

The structure of this paper is a follows: First, we describe the theoretical basis for our intervention in the values in science literature and Rest's theory of moral development. We then describe the value-focused fellowship and provide details of how we analyzed the interviews for moral sensitivity and moral reasoning. We follow with results on how the fellowship changed the scientists' moral sensitivity and their moral reasoning. Finally, we end with reflections on how  our measures worked and ways to improve these measures.

\section{Literature Review}

\subsection{\label{sec:level2}Values in Science}

Science cannot be ``value free". In a recent study, Elliot noted that decisions in science are \textit{necessarily} made with values in mind \cite{elliott_values_2022}. Values are involved in science when choosing research questions, handling data, drawing conclusions, communicating findings, etc. For example, climate modelers have to consider values on the local weather versus the global weather to optimize their model to predict the weather. Scientists conducting vaccine trials must make tradeoffs between production speed, safety for human participants, and the reliability of the efficacy data.

One of the main arguments that science should remain value free is that people will not trust science if scientists place their own personal values in their research \cite{elliott_values_2022,bright_du_2018}. While this might be an ideal worth striving for, it is simply not possible to conduct science without making explicit and implicit value judgements. Elliot mentions in their paper that there are multiple arguments on why science can never really be value free with the gap argument (evidential gaps in data are filled by value-laden background assumptions), the error argument (when scientists face epistemic risks they ought to factor non-epistemic values into decisions), the aims argument (achieving a non-epistemic aim in science means taking into account non-epistemic values when using models, theories, and hypothesis), and the conceptual argument (non-epistemic values are relevant to assessing value-laden concepts) \cite{elliott_values_2022}.

Findings from the developmental moral psychology literature show that changes in moral reasoning are correlated with improvements in ethical conduct in the profession \cite{rest_dit2_1999}. Currently most ethics training focuses on applications of ethical theories or codes of ethics, instead of discussing the values embedded in ordinary scientific practice \cite{mumford_read_nodate}. While scientists receive training in methodologies for distinguishing factual or theoretical questions, they receive little training in how to adjudicate questions about values such as whether to replicate data through a third party or spend more money training a graduate student. 

In a project looking at the values expressed by scientists' when discussing decisions within their own work, Linville et al. found that scientists view the value judgements in science as involving  more than just ethical and epistemic values \cite{linville_how_2023}. For example, when tasked with redoing a study or publishing the study with omitted data, the scientists considered possible funding (economic values) problems with redoing the study. The conclusions for this paper were that ethical training must focus more on the overall values in science rather than focusing on ethical theories or guidelines.

\subsection{Assessing Ethical Decision Making}


Historically someone's ethical decision making is measured through standardized assessments \cite{you_independence_2013}. Despite these assessments being accessible to everyone, these assessments have been studied in the medical/dentistry field with very few studies measuring scientists' ethical decision making.

There are a few assessments that gauge someone's ethical sensitivity. For example, there is the Dental Ethical Sensitivity Test (DEST) and Hebert et al's medical ethical sensitivity test \cite{bebeau_measuring_1985,hebert_measuring_1992}. In both of these measures, the person taking the test reads vignettes regarding ethical scenarios and then is asked to identify the moral issues in each vignette. 


To gauge someone's moral reasoning, moral psychologists have developed the Defining Issues Test (DIT) and subsequently developed the DIT-2 \cite{mumford_read_nodate, rest_dit2_1999}. The DIT-2 is the accepted test to measure moral reasoning, which contains 5 moral dilemmas including ones like stealing food from someone hoarding food and a doctor has to decide to give a frail and suffering patient an overdose on medicine \cite{About_the_DIT}. Currently, participants in other moral reasoning studies are only given the DIT-2 where they are given a multiple choice format to rank a set of items based on the scenarios \cite{rest_neo-kohlbergian_1999}. Another way to measure ethical reasoning capacity is to use the Ethical Decision Making Measure (EDM). This assessment is similar to the DIT where the participant is given summaries of ethical scenarios and asked to rate the severity of the violation in the scenario on a scale of 1 (low) to 7 (high) \cite{mumford_articles_2006}.

One example of a real life application of measuring moral reasoning is when Klinker and Hackmann used a modified version of the DIT-2 to measure school principal's moral reasoning. The study looked at 64 different state school principals with different years of employment to see their level of reasoning when asked about scenarios that could happen to them at their schools. Their reasoning was gauged through a multiple choice test where the participants could pick both their responses and their justifications for those responses \cite{klinker_analysis_2004}. Klinker and Hackmann found that the principals' values is one of the main factors that they use in order to make ethical decisions regarding students' welfare.


Moral motivation has received less attention from the psychology literature than the other components. You and Bebeau measured moral motivation through the Professional Role Orientation Inventory (PROI). The PROI assesses a person's commitment to prioritize professional values over personal values by measuring their authority/responsibility and a person's perceptions of self-efficacy by measuring their agency/autonomy \cite{bebeau_development_1993, you_independence_2013}. 


Moral implementation is the least investigated component. The sole attempt at measurement was by You and Bebeau, who used dental students' scores in a professional problem solving (PPS) class. In the class students were required to implement action for complex cases that present difficult human interaction problems that can arise in dental practice \cite{you_independence_2013}.

\subsection{Approaches to Ethics Training}

One of the main ways for scientists to receive ethical training is through self-paced online modules like the CITI program \cite{mumford_articles_2006}. There are also hybrid options available, but a recent study of university RCR training requirements found that 82\% of research universities with available institutional plans for RCR training require an online-only ethics training \cite{phillips_america_2018}. This training has been found to be ineffective improving ethical decision-making, likely because the online training is wholly didactic, and focuses on applying regulatory rules to simple, context-free cases \cite{antes_meta-analysis_2009}. Current RCR training is often restricted to passive online readings or lectures while it has been found that active learning modalities are more effective than passive learning modalities \cite{dubois_teaching_2009, kalichman_responding_2007}. While it was found that longer ethical training did cause people to have better ethical decision making, many institutions do not have the resources to have these extended training sessions \cite{mumford_read_nodate}.

Some studies have sought to change RCR training to be more grounded in actual scientific practice instead of passively focusing on the rules and regulations of RCR. One of these new methods is to have a virtue-based approach to teaching ethics. This kind of training involves looking at the ethical considerations scientists' encounter on a daily basis and connects RCR to this everyday ethical decision making \cite{pennock_developing_2017}. This would allow scientists to see the connection between ethical values and the practice of science, hopefully leading to more robust and consistent ethical conduct.

One way of changing RCR training to be more grounded in scientific practice is to see the causes of unethical conduct first. Cairns et al. conducted a phenomenography where scientists were asked what makes other scientists act unethical. They found that the two biggest perceived factors for unethical conduct are pressure and personal gain \cite{cairns_phenomenographic_2021}. These two factors they found line up with what other literature says about what motivates people to act unethically.

\section{Theory}

\subsection{Values}

There are multiple definitions of the concept of ``value". For this study, we adopt the view that an individual values something, if the individual \textit{prefers} that thing over some other thing \cite{hausman_2011}. Values can be ``goals", ``aims", ``principles" or anything else individuals want to be realized. Things can be valued intrinsically, because they are preferred on their own, or instrumentally, because they help individuals achieve other goals. Values that might guide scientific decision-making include finding the truth, getting a publishable result, saving time or money, helping graduate students, providing benefit to society, or a variety of other personal, professional, epistemic, or ethical goals. In general, we distinguish ``values" from ``beliefs" by suggesting that the latter involves claims about purely factual matters (i.e. what \textit{is} the case), while the former involves claims about normative ideals (i.e. what \textit{ought} to be the case). Note that values can be held by a community as well as individuals, and they need not be mere ``opinion" or ``taste". Ethical values are values, and at least some people hold some ethical values to apply universally \cite{kagan_normative_2018}.

Our main focus in this paper is on the different values scientists invoke when discussing ethics in science. While we should generally expect that scientists will be attentive to aspects of their work relevant to knowledge production, their ability to reason well depends also on their attentiveness to ethical values. Therefore, we investigated the \textit{epistemic}, \textit{ethical}, and  \textit{other} non-epistemic values invoked by scientists during discussions about research ethics. In this research we relied on the ``values in science" literature that notes the many ways decision-making about scientific methods are laden with non-epistemic (e.g. ethical, legal, or economic) values \cite{brown_values_2013, douglas_inductive_2000}.

\subsection{Rest's theory of moral development}

The Four Component Model (FCM) states that in order for someone to behave morally in a situation, they must demonstrate each of the following components: moral sensitivity, moral reasoning, moral motivation, and moral implementation \cite{rest_moral_1994}. These components are not steps that have to be followed in a linear sequence but are rather mental processes that collectively and interactively produce ethical action \cite{rest_research_1984}. A summary of the four components for ethical decision making are found in Table \ref{tab:table1}.

\begin{table}[]
\begin{ruledtabular}
\begin{tabular}{p{3.2cm}  p{5.3cm}}
\textbf{Component}   & \textbf{Description}          
 \\ \hline 
Moral Sensitivity    & Identifying morally salient ideas    \\
Moral Reasoning      & Identifying morally right action     \\
Moral Motivation     & Intending to do morally right action \\
Moral Implementation & Acting on morally right action      
\end{tabular}
\end{ruledtabular} 

\caption{\label{tab:table1}
Brief definitions of the four components of ethical decision-making}
\end{table}

Moral sensitivity is the ability to identify the morally salient features of the scenario. This component is where the person making an ethical decision identifies what is going on in a scenario and how a decision might affect others \cite{rest_moral_1994}. For example when scientists want to reproduce their results for a study, they must evaluate the ethical dimensions of different alternatives: training grad students to reproduce the data or send the data to a third party to reproduce the data and save money. 

Moral reasoning is the capacity to identify the morally right action (e.g: identifying the option to benefit society instead of the option to benefit oneself) \cite{rest_research_1984}. This component interacts with moral sensitivity since once someone has identified the morally relevant factors of the situation (people possibly affected, actions available, etc), they must decide which action or actions are morally justifiable \cite{rest_moral_1994}. Rest et al. developed a set of "Neo-Kohlberg" schemas to classify moral reasoning based on Kohlberg's theory of moral development \cite{thoma_defining_nodate}. Rest et. al. organized moral reasoning into three schemas (1) personal interest, (2) maintaining norms, and (3) post-conventional \cite{thoma_defining_nodate}. 

The personal interest schema focuses on the personal gains and losses that result from moral action. The maintaining norms schema focuses on following some set of rules or guidelines from moral action \cite{rest_postconventional_1999}. The maintaining norms schema contains the following qualities: generally accepted social norms as a collective, norms apply society-wide, need for norms to be clear, norms establish a reciprocity, and there is an establishment of a hierarchy \cite{thoma_measuring_2014}. The post-conventional schema focuses on prioritizing what society idealizes (e.g: maximizing other people's welfare) \cite{rest_postconventional_1999}. The post-conventional schema contains the following qualities: involves reasoning about moral obligations based on shared ideals, fully reciprocal, and are open to scrutiny \cite{rest_dit2_1999,thoma_measuring_2014}. These three schema are summarized in Table \ref{tab:table2}. In Rest's formulation of the DIT-2 there is a hierarchy between the schema; The DIT-2 is scored based on someone's N2 score which measures how much someone prioritizes the post-conventional schema over the personal interest schema \cite{About_the_DIT}, which is determined by the degree to which post-conventional items are prioritized and personal interest items are deprioritized. 

Moral motivation is the intent to do what is morally right \cite{rest_research_1984}. After someone identifies the morally justifiable decision, they must want to act on this decision. This means that someone does not always need moral reasoning or moral sensitivity to have moral motivation. Rest argues that in order for someone to have a high moral motivation, they must prioritize ethical values over other values \cite{rest_moral_1994}. Someone with a low moral motivation might be able to identify the ethical values and reasoning, but they might not prioritize these values which will not lead to a morally justifiable action.

Moral implementation is where the person acts on their plan of action. A person may be able to identify an action, come up with a morally justifiable reason for the action, and place a priority on moral values, but in the end they could end up not doing any action because of a low moral implementation (i.e. non-cognitive barriers to action such as weakness of will, lack of time, anxiety, post-traumatic stress, etc.) \cite{rest_moral_1994}.

While these components can interact with each other, these components are distinct from one another, meaning that mastery in one of the components does not mean mastery in other components. You and Bebeau found that after testing students on their proficiency in each of the four components, there was no correlation between a student's competence in one component compared to another \cite{you_independence_2013}. 


\begin{table}[h]
\begin{ruledtabular}
\begin{tabular}{  p{2.7cm}  p{5.5cm}  } 

Neo-Kohlberg Schema & Definition  \\ 

  \hline
 Personal Interest Schema & Doing something in hopes of personal gain \\
  Maintaining Norms Schema & Doing something to follow set guidelines (e.g. following the law, following set science standards)   \\ 
  Post-Conventional Schema & Doing something because it is ultimately the right thing to do  \\
 
\end{tabular}
\end{ruledtabular} 

\caption{\label{tab:table2}
Brief definitions of the schema used in moral reasoning}

\end{table}

\subsection{Improving Ethical Decision Making}

The main reason for using the FCM is that it provides a schematic for identifying the components in a person's ethical decision-making. While we can identify values to identify the connection someone makes between ethics and science, we need to also use the FCM to identify how someone makes a decision in science. Identifying a person's components of their ethical decision will allow us to focus research and interventions to strengthen any deficiencies in any of the components. For example, some research has identified that an improvement in someone's moral sensitivity will lead to an overall improvement in ethical decision making, indicating that training should focus primarily on improving someone's moral sensitivity \cite{myyry_role_2002,clarkeburn_impact_2002}.

We argue that someone needs to improve on all of the ethical decision making components in order to improve on ethical decision making. Narvaez and Rest claim that a person who fails to make the morally right decision has a deficiency in one or more of the ethical decision making components \cite{narvaez_four_1995}. For example, someone can identify all of the morally salient ideas in a scenario but ultimately cannot come up with a reason to make a decision. Since all components are a necessary part of the process for making an ethical decision, an improvement in one component \textit{does not guarantee} an improvement in someone's ethical decision making \cite{narvaez_four_1995}. We then define improving in ethical decision making as improving in all of the components, but an improvement in one or more of the components is a step in the right direction towards morally acceptable decisions.

We identify how scientists' are improving with respect to individual ethical decision making components. As mentioned in the introduction, we focused on changes in moral sensitivity and moral reasoning. Since Rest defined moral sensitivity as identifying the morally salient ideas in a scenario and the literature states that epistemic and non-epistemic values are important for making decisions in science \cite{brown_values_2013,elliott_values_2022,douglas_inductive_2000}. 

We define an improvement in scientists' moral sensitivity as invoking a greater number of values when discussing ethically-laden cases. ``More values" in this case means quantitatively more values and more types values (e.g. instead of invoking purely epistemic values, participants should invoke ethical, epistemic, and other values in their sensitivity).

We define an improvement in moral reasoning as reasoning with a higher level schema. As mentioned above, in the DIT-2, the post-conventional schema is defined the  as the highest level reasoning schema, followed by the conventional and the personal interest schemas \cite{About_the_DIT}. For example, someone's initial reasoning for a scenario starts in the self-interest schema but over time the reasoning changes into the post-conventional schema. This means that their moral reasoning improved. In Rest's case he applied these schema to multiple choice questions. In this study we applied the same schema to participants' open-ended responses to ethical vignettes.

We applied these theories to answer the following research questions: 
\begin{enumerate}
  \item How do discussions of values embedded in science affect scientists' \textit{moral sensitivity}?
  
  \item How do discussions of values embedded in science impact scientists' \textit{moral reasoning}?
  
  \item How effective is the value-focused training in improving scientists' overall ethical decision making?
\end{enumerate}

\section{Methods}
\subsection{Fellowship}

Our project is structured around interviews conducted as part of a year-long fellowship originally for 15 faculty members in the basic sciences at a midwestern research university. One of the faculty members dropped out so we had 14 faculty members for this study. Participants were recruited by email, word of mouth, and explicit invitations by the fellowship organizers. Recruitment intentionally tried to promote diversity in terms of gender and academic status. 

The fellowship consisted of 10 sessions over the course of an academic year. Each fellowship session was one and a half hours long, and centered on reading materials given to participants prior to the session. The fellowship did not explicitly teach the participants about ethics, but was designed to foster directed exploration and discussion of the role that values play in decisions made within science. The contexts ranged from foundational methodological decisions, such as the choice of statistical methods, to decisions about the application of science, such as whether and how to advocate for policy changes. Throughout the fellowship, the participants were made aware that they were explicitly addressing value questions. They knew the context of the research project was about ethics in science, but the research goals were not discussed.

Author Scott Tanona ran the sessions and helped guide discussions. Each session focused on a particular topic meant to highlight the role of values in scientific decision-making. The sessions were designed to provide some basic conceptual tools for understanding and potentially addressing issues. Participants were asked to read and come prepared to discuss pre-readings they were provided for each session. During the sessions, the participants were prompted to discuss topics with open-ended, general questions inviting them to react or discuss what they thought was interesting, as well as with directed questions meant to elicit discussion of particular topics. At the end of the session, they were asked to write a follow-up reflection, which asked for general thoughts about the session as well as for answers to specific follow-up questions. The 15 (eventually 14) fellowship participants were split into two even groups for ease of scheduling and to ensure more chances of discussion from each participant. Halfway through the fellowship (after five sessions), the groups switched some members for scheduling purposes and to provide broader exposure to other faculty. Meetings were held in person, until the COVID-19 pandemic required meetings by Zoom for the last three sessions.

The first introductory session introduced participants to ethical theories and other philosophical concepts (such as descriptive and normative distinction). After that, each session focused on a particular topic or two that highlighted the potential role of values in scientific decision-making, including a number of topics from the ``values in science" literature (e.g., inductive risk, choice of statistical methods, social structure of science, reproducibility) and topics that highlighted the connection between science and society (e.g., advocacy, diversity, public participation in science, public sharing of pre-prints). In the final sessions, discussion turned to the COVID-19 pandemic, during which previous concepts were applied.

Topics were chosen and sessions were designed to encourage participants to reflect on and communicate both the immediate, intermediate, and long-term instrumental goals of their own work. These topics highlighted epistemic values (including choices between them, e.g., between avoiding type I or type II error, or between modeling for predictive accuracy or representational accuracy), as well as relationships between epistemic goals and ethical values (e.g., in ``inductive risk"). This allowed discussions about the goals of science as a whole, disciplinary standards, general disciplinary goals, and impacts on society. This same fellowship is explored in other works \cite{cairns_phenomenographic_2021, linville_how_2023}.

\subsection{Data Collection - Interviews}

Our data for this project comes from 28 individual, semi-structured interviews of fellowship participants conducted both before and after the fellowship series, respectively called the pre-interview and the post-interview. All interviews were video and audio recorded. Due to one of the fellowship participants dropping out and being unable to make time for a post-interview, we only analyzed the pre- and post- interviews for 14 of the 15 participants. In these interviews, the interviewee was asked to respond to a series of fictional vignettes involving research misconduct (named “RCR Vignettes”).  In these recordings, the interviewer asked follow questions to the interviewee's responses to the vignettes that were supposed to elicit more detailed responses from the interviewee (e.g. ``Can you explain more about ..."). In total, there are three vignettes with six different ethical dilemmas within these vignettes. Two of the three vignettes are adapted from the Ethical Decision Making-measures assessment (EDM)\cite{mumford_articles_2006}. The names and brief descriptions of the dilemmas can be found in Table \ref{Vignette}.  Note that while these vignettes are not specifically related to physics, these concerns are relevant to physicists \cite{kirby_ethics_2004}.

We transcribed all 28 (14 pre-interview plus 14 post-interview) of the video and audio recordings using otter.ai as the initial transcriber. We then corrected the responses starting from the first question of the interview and ended as soon as the questions about vignettes were done.

\begin{table}
\begin{center}
\begin{ruledtabular}
\begin{tabular}{p{2cm} p{6.3cm}} 

  \textbf{Title} & \textbf{Description} \\ 
  \hline
  Informed consent & A professor wants to conduct an experiment where he gives a placebo shock to participants, but does not tell the participants about the placebo. The graduate student working on the project does not want this project to go on. \\ 
  Publishing concern & The previous experiment ends up being conducted and two groups are working on the data. One group believes that their data is incorrect, and both groups want to publish the correct data.  \\ 
  Authorship concern & The professor promises a graduate student that they would be first author on the project, but he then takes back this promise because he needs more first authorship.  \\
  Review similar experiment & A professor is asked to review a project, but the project is very similar to the one he is working on. \\
 Unpublished data & The professor found out that one of his students got a hold of unpublished data from another student and is trying to use the data in their own research. \\
  Hiring diversity & A professor department is hiring a new faculty member. They can only bring in three candidates, and they decide to bring in three male candidates and leave out a female candidate.\\
  
\end{tabular}
\caption{\label{Vignette}
Brief description of the ethical dilemmas from the interviews. }
\end{ruledtabular}

\end{center}
\end{table}

\subsection{Data Analysis}

In order to see what \emph{kinds} of values scientists appealed to when thinking about research ethics, we identified the explicit and implicit content of the values they appealed to when discussing the vignettes. We are primarily interested in the extent to which scientists invoke epistemic, ethical or other non-epistemic and non-ethical values. We first took all of the responses from the interviewee and split them up into quotes. One quote was the part of the response that the interviewee invoked a value or values.

We identified values from quotes by identifying expressions that contained normative or evaluative judgements. A feature of the statement was judged as valued in a statement if it could be fit into a form such as ``The interviewee values..." or ``the interviewee has a goal of ...", whether or not the person explicitly held that value themselves or was discussing someone else's values or possible values. An example of how we coded values from quotes is found in the Moral Sensitivity section.

We identified improvement in moral sensitivity by looking at \textit{counts} of the values invoked. For an improvement in moral reasoning we looked at \textit{percent distribution}. Due to the independence of improvement on the components (an improvement on one of the components does not mean an improvement on another component), we accept that we can analyze the two components for improvements in different ways. 


\begin{table*}
\caption{\label{Values}
This table shows the categories, sub-categories, and definitions for the values used in the project. This table is inspired by Linville's values table \cite{linville_how_2023} }
\begin{ruledtabular}
\begin{tabular}{p{3cm} p{3cm} p{11cm}} 
  Category & Subcategory & Definition \\ 
  \hline
  Ethical & Ethical, misc. & Doing what is right. This subcategory was used when the interviewee valued doing what is right, but it did not fit in the other subcategories \\ 
   & Rights & Treating people in a certain way due to some intrinsic feature \\ 
   & Fairness & Treating others in the same way \\
   & Social Good & Doing something that will benefit society or a group of people \\
 & Virtue & Fulfilling a character trait that the interviewee found desirable \\
  & Interpersonal Care & Maximizing welfare for one specific person\\
  Responsible Conduct of Reasoning (RCR) & RCR & Favoring regulations set by RCR training (plagiarism, authorship, etc) \\
  Legal & Legal & Worrying about threats by government regulations.\\
  Communitarian & Communitarian & Doing something good for the desire for peer/social approval. \\
  Epistemic & Epistemic, misc & Advancing or improving science in some way. This subcategory was used when the interviewee valued improving science, but it did not fit in the other subcategories \\
  & Alethic & Pursuing or clarifying knowledge about something \\ 
   & Explanatory & Understanding some process \\
   & Methodological & Understanding a part of the scientific process \\ 
   & Aesthetic & Having a theory or explanation that is pleasing to the interviewee \\
   & Predictive & Conducting science to make predictions about the future \\
   & Empirical & Disseminating data or proper use of data \\
   & Technological & Using science to have some technological application \\
  Economic & Economic & Using fixed resources (time, money, etc) \\
  Self-Interest & Self-Interest & Action providing some benefit to the interviewee \\
  Practical & Practical & Actions that must be done for science to be conducted \\
  
\end{tabular}

\end{ruledtabular}
\end{table*}

\subsubsection{Moral Sensitivity}

As mentioned in the theory section, we define an improvement in moral sensitivity as participants invoking more epistemic and non-epistemic values. We focused on changes in scientist's ``moral sensitivity" by looking at the different values the scientists invoked since these are the ``morally salient features" of the scenarios that we gave them. 

Originally we defined the types of values scientists can invoke in ethical scenarios by using Kohlberg's stages of moral development. These values are ethical (appeal to what is the right thing to do), legal (appeal to threat or punishment by governing entity), communitarian (appeal to social approval), and self interest (appeal that benefits self) \cite{kohlberg_moral_1983}. As we coded the quotes, we found other emergent types of values. These values are epistemic (appeal to improving knowledge), RCR (appeal to regulations and guidelines), economic (appeal on using resources), and practical (appeal to actions that must be done for science). Within the ethical and epistemic categories are sub-categories that further defined the categories of these values. The definitions for the categories and sub-categories are in Table \ref{Values}.


We also investigated how participants invoked \textit{unique} values. We define unique values as distinct and separate values that are not mentioned previously in the scenario. We are identifying unique values since we are saying that Moral Sensitivity is being able to identify the morally salient features in a scenario, we looked to see if the participants are invoking a wide range of values instead of repeating the same value over again. We started by looking at the quotes to see the sub-category of the value for the quote. If there are two values with the same sub-category within the same scenario, then we would return to the original transcript and look at the context for the value. For example, if someone values ``taking care of their students" followed by valuing ``publishing accurate data" and then said they valued ``taking care of their students" in the same scenario, there would be three total values since they are valuing three things in total, but there would only be two unique values since they are valuing ``taking care of their students" twice. 

For this analysis, we considered the unique values \textit{per person}, so if one of the participants said that they value ``taking care of their students" and another participant said that they also value ``taking care of their students", then these would be considered two different unique values when looking at the data since these values come from two different people. Below are example quotes for coding values:

\textbf{Ethical Value Example:} In this quote the interviewee is responding to the question ``What should Dr. Kaylee do?" about trying to convince other people to bring in a woman to the hiring process when discussing the ``Hiring diversity" dilemma:

\begin{quote}\vspace{-.3\baselineskip}
    {\em ``So I mean, I guess if they're all roughly equal I mean he should probably talk to the chair and get the woman in as a candidate...\textbf{It's kind of really important to have the same sort of representation as the general population}. It's gonna be problematic if you don't have that.''} - Interviewee 2
\end{quote}\vspace{-.3\baselineskip}

In the interviewee's response it shows that they think that having a diverse representation in the department is ``important" and that the department should bring in the woman candidate. This shows that they value the importance in having proper and equal representation in a department and since this would be ultimately benefiting other people in the department this would be an ethical value with the subcategory of ``fairness".

\textbf{Epistemic Value Example:} This quote the interviewee is responding to the interviewer's statement ``Tell me why you shouldn't throw out data" when following up on the question ``What should they do?" regarding the discrepancy between the two group's data:
\begin{quote}\vspace{-.3\baselineskip}
    {\em ``So having data sets that \textbf{acknowledge all of your data, how it was collected, and then that it's been thoroughly vetted} so that you've done your best effort possible to make sure that there haven't just been mistakes in the synthesis of these data sets. \textbf{I think it's integral to getting that data}.''} -Interviewee 14
\end{quote}\vspace{-.3\baselineskip}

The interviewee responded by saying that all data and the collection and analysis of the data is ``integral" and should not be thrown out. This quote is coded as ``epistemic" with the subcategory of ``empirical" as we see that the interviewee values having proper collection/analysis of all data.

We identify the difference in the \textit{counts} of values between the pre-interviews and post-interviews. We did not assume any hierarchy among the values, and thus focused on how many values are invoked and how many values of each category are invoked. An analysis of invoking epistemic/ethical values versus RCR values can be found in Linville et. al. \cite{linville_how_2023}.

We used a paired t-test to determine if there are any significant differences between the pre-interview values invoked and post-interview values invoked. This test determines if there is a mean difference between observations for two pairs of data (in our case the pairs would be the pre and post values) \cite{hsu_paired_2008}. Due to the low numbers of scientists, we added up all of the scientists' invoked values per ethical dilemma for the pre and the post interviews. We then ran a paired t-test on the total values invoked per dilemma and unique values invoked per dilemma to find the p-value for the differences in values invoked. 


We also found Cohen's d to further determine if there are any significant differences in the values invoked. Cohen's d determines our standardized effect size of our data, which allows us to see the variation of the means of values invoked between the pre-interviews and post-interviews \cite{lakens_calculating_2013}. We then ran a program in R using the ``effsize" library that found the Cohen's d between the pre and post-interview values. The generally accepted value of Cohen's d are as follows: d = 0.2 is a small effect size, d = 0.5 is a medium effect size, and d = 0.8 is a large effect size \cite{lakens_calculating_2013}.

\subsubsection{Moral Reasoning}

We focused on changes in scientist's ``moral reasoning" by identifying the participants' Neo-Kohlberg Schema in their reasoning to questions about the vignettes. At first, we coded each individual response. Identifying an individual response is defined as a claim made by the participants followed by the reasoning for that claim statement. Note that this order of claim then reasoning could be flipped and we needed both the claim and reasoning to be considered an individual response. 

 An example of identifying a response is found in the quote ``they should stop the experiment because harming people is not the right thing to do”. In this quote, the claim made is that ``they should stop the experiment". The reasoning following the claim is ``harming people is not the right thing to do". We only applied the Neo-Kohlberg Schema to the participants' reasoning and not the claims they were making.

At first, we coded each individual response. After coding each of the individual participant's responses, we combined all of the interviewee responses to see how the scientists’ moral reasoning as a whole changed. We identified the participants' moral reasoning this way as we are attempting to find a way of identifying moral reasoning in a more asset based model, which the DIT and EDM do not fit in. Below are examples of coding participant's moral reasoning:

\textbf{Post-Conventional Reasoning Example:} In this quote the interviewee is responding to the question ``What should Fowler do?" about being first author on the paper instead of his graduate student:

\begin{quote}\vspace{-.3\baselineskip}
    {\em ``In my opinion, this is one of the most important things we do is build up our students and giving them opportunities to grow and succeed and to develop a CVs that will help them be successful long term is just the highest importance to me.''} - Interviewee 4
\end{quote}\vspace{-.3\baselineskip}

In this quote we classify the claim as ``building up students". The \textbf{reasoning} is the ``being successful in the long term" they get from being first author. The interviewee believes that helping graduate students with their careers is the most important idea when it comes to who's first author and is an example of post-conventional reasoning (helping others, ultimately right thing to do).

\textbf{Maintaining Norms Reasoning Example:} In this quote the interviewee is responding to the question ``What should Stavenick do?" regarding a grad student using unpublished work in their research:

\begin{quote}\vspace{-.3\baselineskip}
    {\em ``So I mean, that would be a discussion along the lines of you, you know, you can't do this because you're not allowed to see this proposal in the first place.''} -Interviewee 12
\end{quote}\vspace{-.3\baselineskip}

The claim in the quote is ``[the student] can't do this" study. The \textbf{reasoning} is the grad student is not allowed to see the proposal in the first place. The use of ``not allowed" shows that the interviewee is referencing rules that prevent the grad student from seeing the unpublished data in the first place and is an example of maintaining norms reasoning (following a set of rules set in the science community).

Since we are identifying an improvement in moral reasoning as participants using higher level schema in their reasoning after the fellowship, we analyzed the participants' \textit{percent distribution} of schemas to determine if the participants' reasoning changed. Since we are now looking at which specific schemas the participants are using due to the hierarchy of the schemas, a percentage breakdown of each schema the participants use will more accurately reflect on their change in moral reasoning. We do identify the counts of schema as a way to see how the number of reasonings are effected by the number of values.

We used Fisher's exact test to compare the distributions of schema between pre-interviews and post-interviews \cite{bewick_statistics_2004}. Fisher's exact test is used to compare the distribution of a categorical variable in a group with the same variable of another group \cite{kim_statistical_2017}. In our case, the categorical variable is the schema invoked and the two groups we're comparing are pre-interviews and the post-interviews. If the Fisher's exact test shows that there is a significant difference in the distribution of schemas between the pre- and post-interviews, then we can say that improvement was a result of the fellowship.

We identified any significant changes in the total number of schemas when we ran a paired t-test to find the p-value for the differences in the schemas invoked. If the p-value is less than 0.05, we will consider this differences significant and conclude that the number of reasonings between the interviews are different. 

\subsubsection{Inter-rater reliability}

We validated the values and Neo-Kohlberg schema through inter-rater reliability (IRR). Three people coded the interviewee's responses. One of the coders coded all of the quotes for both values and responses for the Neo-Kohlberg schema and is labeled the \textit{original coder}. The other coders received 120 quotes (around 20\% of the total quotes) to determine the values and 84 responses (around 20\% of the total responses) for schema. 

We used the Fleiss Kappa to determine the rater agreement between the three researchers for the individual values \cite{fleiss_measuring_1971}. Calculating the Fleiss Kappa gives a number between 1.0 and 0.0 where 1.0 means the coders had a 100\% agreement between codes and 0.0 means the coders had a 0\% agreement between codes. While the Fleiss Kappa value is calculated by counting the agreement between the coders, there are many outside factors affecting value like how long people have been in the project and how many categories of values that are coded \cite{hallgren_computing_2012,sim_kappa_2005}. 

For coding the values in general, the original coder sent out around 20\% of the total value quotes coded (around 121 quotes) and the other two researchers coded values for these quotes. The Fleiss Kappa value for the reliability is around 0.53, which according to Landis and Koch would be a moderate agreement between the coders \cite{landis_measurement_1977}.  After the Fleiss Kappa value was calculated, the two researchers met up with the original coder and discussed the values they coded. After discussing the values the coders agreed on the same value for each quote until there was a 100\% agreement between the coders for the values for each quote that was sent to the other coders. 

For the unique values, the coders went through each ethical dilemma and discussed what they thought were unique values. We looked for agreements/disagreements by discussing what were the unique values in each dilemma. After going through one of the interviewee dilemmas that were sent out and the coders reached 100\% agreement on each of the unique values, the coders continued to the next one until 100\% agreement was reached for each unique value on the dilemmas that were sent to the other coders.

For the IRR for the moral reasoning schema, we used a similar method to the total number of values where the original coder sent out around 20\% of the total reasoning responses and the other two researchers coded schema for these responses. The Fleiss Kappa value for the reliability is around 0.39, which according to Landis and Koch would be a fair agreement between the coders \cite{landis_measurement_1977}. After the Fleiss Kappa value was calculated, the two researchers met up with the original coder and discussed the schema they coded. The coders then discussed the agreed upon schema for a response. After discussion there was a 100\% agreement between the coders for each schema for the send out responses.

\subsection{Limitations}

We have a relatively small number of faculty from a single university in this study. This allows us to look at a small section of the scientific community and we therefore hesitate to make broad claims from this. Additionally, we acknowledge that the values participants' invoked could have changed due to factors outside the fellowship.

We also acknowledge that due to the one-on-one nature of the interviews, the participants may have displayed some sort of social approval bias where they mainly talked about doing what is right. This may raise the total post-conventional reasonings, but likely did so in both the pre-interviews and post-interviews.

\section{Results}

We found that the average interview length for the vignettes was about thirty minutes for both the pre-interviews and the post-interviews (the average interview length for the pre-interviews was 29:41 while the average interview length for the post-interviews was 30:30)

\subsection{Moral Sensitivity}

\begin{figure}[th]
    \centering
    \includegraphics[scale = 0.5]{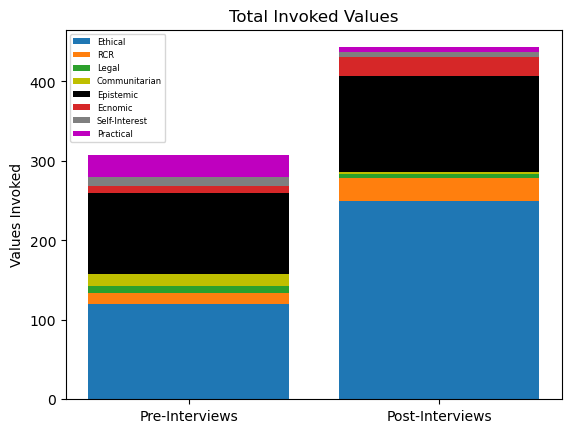}
    \caption{Total invoked values for the pre-interviews and post-interviews. We combined all participants into one graph since we are looking at scientists as a whole.}
    \label{Total_Values}
\end{figure}

Figure \ref{Total_Values} shows that the total number of values invoked increased in the post-interview (443 vs 308). The most prevalent values to show up are the ethical values and the epistemic values. While these two categories show up the most, there is only a change between the pre-interviews and the post-interviews for the ethical category. The ethical category had the largest increase  in number of values invoked while every other category stayed roughly the same.

We found that the paired t-test for the total values gives us a p-value of 0.007 and a Cohen's d of 1.13 (large effect size). The difference between the post-interview values invoked and pre-interview values invoked are statistically significant and substantial. More explicitly, the number of post-interview values invoked are significantly higher than the pre-interview values invoked.

\begin{table}[]
\begin{center}
\begin{ruledtabular}
\begin{tabular}{l r r}
\textbf{Ethical Dilemma}      & \textbf{Pre Values} & \textbf{Post Values} \\
Informed Consent       & 67                          & 92                           \\
Publishing Concern     & 73                          & 107                          \\
Authorship Concern     & 36                          & 42                           \\
Review Similar Project & 48                          & 74                           \\
Unpublished Data       & 33                          & 68                           \\
Hiring Diversity       & 51                          & 60                          
\end{tabular}
\end{ruledtabular}
\caption{\label{Total Values}
Total values invoked per ethical dilemma from the interviews. This data was used to run the t-test and to find Cohen's d}
\end{center}
\end{table}

\begin{figure}[th]
    \centering
    \includegraphics[scale = 0.5]{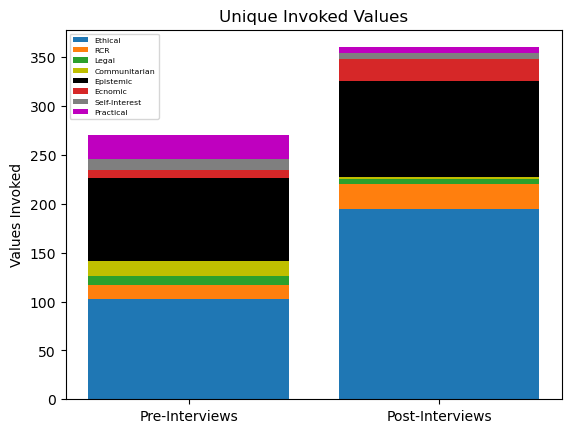}
    \caption{Chart looking at the difference between unique invoked values for the pre-interviews and post-interviews. We combined all participants into one graph since we are looking at scientists as a whole.}
    \label{Unique_Values}
\end{figure}

 Looking at unique values invoked in Figure \ref{Unique_Values}, we see that the participants are invoking a variety of values instead of repeating the same values. The trends here are similar to what we see when looking at the total number of values invoked. There is an increase in the total number of unique values invoked (360 vs 270). Once again, the biggest increase in unique values was in the ethics category while the rest of the categories stayed relatively the same.

 We found that the paired t-test for the unique values gives us a p-value of 0.02 and a Cohen's d of 0.97 (large effect size). There is a significant difference between the post-interview unique values and pre-interview unique values.

\begin{table}[]
\begin{center}
\begin{ruledtabular}
\begin{tabular}{l r r}
\textbf{Ethical Dilemma}      & \textbf{Pre Values} & \textbf{Post Values} \\
Informed Consent       & 60                          & 75                           \\
Publishing Concern     & 58                          & 83                          \\
Authorship Concern     & 32                          & 36                           \\
Review Similar Project & 45                          & 66                           \\
Unpublished Data       & 33                          & 57                           \\
Hiring Diversity       & 42                          & 43                          
\end{tabular}
\end{ruledtabular}
\caption{\label{Unique Values}
Unique values invoked per ethical dilemma from the interviews. This data was used to run the t-test and to find Cohen's d for the unique values}
\end{center}
\end{table}

\subsection{Moral Reasoning}

\begin{figure}[t]
    \centering
    \includegraphics[scale = 0.5]{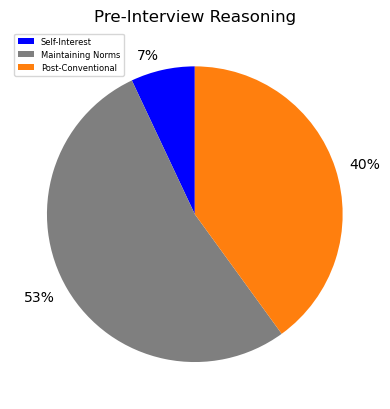}
    \caption{Pie chart for percent breakdown of schema for levels of reasoning for the pre-interviews. We combined all participants into one graph since we are looking at scientists as a whole.}
    \label{Pre_Reasoning}
\end{figure}

\begin{figure}[t]
    \centering
    \includegraphics[scale = 0.5]{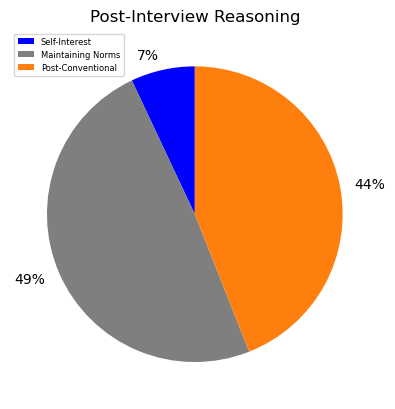}
    \caption{Pie chart for percent breakdown of schema for levels of reasoning for the post-interviews. We combined all participants into one graph since we are looking at scientists as a whole.}
    \label{Post_Reasoning}
\end{figure}

Figure \ref{Pre_Reasoning} and Figure \ref{Post_Reasoning} show there is no significant change in the distribution of schemas between the pre-interviews and the post-interviews for the participants' moral reasoning. The Self-Interest schema did not change with staying at 7\% for both the pre-interviews and post-interviews, while the Maintaining Norms schema and the Post-Conventional schema showed no significant change between the interviews as the post-conventional schema increased by 4\% in the post-interviews.

The Fisher's exact test for the \textit{distribution} of schemas invoked gives us a p-value of 0.70. The difference between the distribution of pre-interview schemas and post-interview schemas is insignificant.

The paired t-test for the \textit{total number} of schemas invoked gives us a p-value of 0.004 and a Cohen's d of 1.34 (large effect size). The difference between the the total post-interview schemas invoked and total pre-interview schemas invoked are statistically significant. More explicitly, the number of post-interview schemas invoked are significantly higher than the pre-interview schemas invoked.

\begin{figure}[t]   
 \centering
    \includegraphics[scale = 0.5]{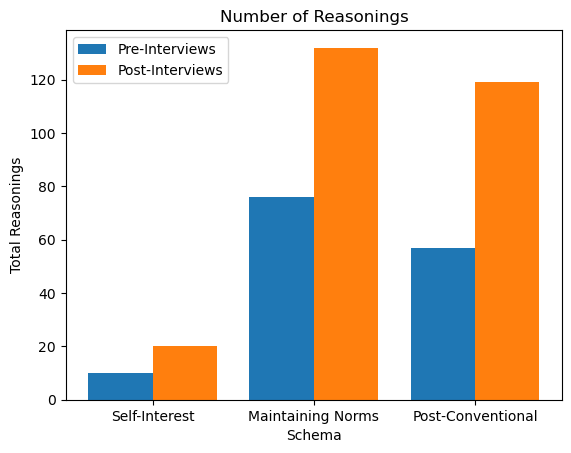}
    \caption{Bar chart for every reasoning categorized into the different schemas for all of the scientists.}
    \label{Total_Reasoning}
\end{figure}

\section{Discussion and Future Work}

To answer the question of ``how do the discussions of values embedded in science shift scientists' ethical sensitivity", we claim that the value-focused training improves scientists' ethical sensitivity. The participants evoked more values in the vignettes since there was a significant increase in the \textit{number} of ethical values invoked after the fellowship. 

The participants invoked a wider range of values after the fellowship which is seen in the unique values invoked. Since time is not a factor in the change in number of values, the participants are noticing more values in the vingettes after the fellowship sessions. Since we are saying that the morally salient items in these scenarios are the values, this means that the participants' moral sensitivity increased. This is a step in the right direction since to start improving in ethical decision making people need to improve on each individual component.

The fellowship did not appear to improve scientists' moral reasoning according to our measure. From the data, there is a significant change in the total number of schemas invoked, but there is not a change in the distribution of schemas participants used. While there is a significant increase in the \textit{number} of distinct reasons brought up from pre to post fellowship (as seen in Figure \ref{Total_Reasoning}), the overall trend between the pre and the post-interviews shows that the schema of reasoning used remains about the same. 

The way we measured the scientists' moral reasoning is not sensitive enough did not fully capture significant changes in moral reasoning. Although the participants' reasoning schema did not change, something about their reasoning must have changed. Reasons for these speculations are because of the \textit{increase} in participants' total reasonings in the post-interviews and the unique values invoked are higher in the post-interviews. This suggests that the participants are able to come up with not only more claims and reasonings for scenarios but they are reasoning with different ideas as well. One suggestion for this is that looking at just the Neo-Kohlberg Schema is not enough to fully analyze someone's moral reasoning as this does not fully encapsulate what is being said in the scientists' moral reasoning \cite{garcia_examining_2022}.

This raises the question ``How do we look more in depth at the participant's reasoning?" Answering this new question will allow us to ultimately answer how the participants' reasoning changed due to the fellowship more accurately. While we  believe that the Neo-Kohlberg Schema should not be disregarded, we need more than just the schema to categorize scientists' reasoning. One part of the next step is to look at how scientist's identify people affected by other people's actions (acting stakeholder and effected stakeholder). We plan to explore this, and other theories from outside of physics, in future projects.

Looking at the effectiveness of the value-focused fellowship on improving the overall ethical decision making of the scientists, we see a start for improving scientists' ethical decision making. We see an improvement in ethical sensitivity and inconclusive results for improvement in moral reasoning. More work will be needed in looking at motivation and implementation in order for us to fully claim that this value-focused training is effective at improving ethical decision making.

One of the goals for this paper is to introduce theories from philosophy and moral psychology to Physics Education Research in order to deconstruct scientists' ethical decision making. We found that one of the better ways to categorize ethical decision making is by using the Four Component Model due to the distinct and easy to recognize components. We believe that as interest in  research ethics education for physicists grows, the Four Component Model (and theories like it) will provide much-needed theoretical and methodological structure to attempts to understand, measure, and improve research ethics education.

Another goal for this paper is to analyze the affects of a novel approach to ethics training by introducing value focused discussions for different science topics. As discussed in Linville et. al's paper, the scientists our study invoked both epistemic and non-epistemic values when discussing ethics \textit{in their work} before the fellowship\cite{linville_how_2023}. If the scientists truly valued RCR training, then we would have seen more invoked RCR values from the participants. Scientists valuing values other than RCR values is further highlighted in this paper as the participants are invoking significantly more ethical values and not RCR values when discussing vignettes instead of their own work after the fellowship. This further highlights the need to improve on current RCR training by focusing more on values in science.

With this work we want to highlight that this kind work required collaboration between physics and philosophy. This is due to the fact that physics in general has little work when it comes to making ethical decisions in science. Because of this, we have to turn to the field of psychology and philosophy to learn how people make ethical decisions and adapt this to science as they are the experts in this field. We encourage future ethics work to be interdisciplinary with philosophy to have a more solid foundation in ethics work. 

\section{Conclusion}

We found that guided deliberations on the values implicit in real examples of scientific practice shifted scientists' invoked values with respect to research ethics. After scientists talked about values at their fellowship meetings, we found that scientists' moral sensitivity improved: they not only noticed more values in the RCR Vignettes, but also invoked more unique values. We did not see an improvement in their moral reasoning though we suspect there was a lack of sensitivity in our way to measure moral reasoning. This study shows that their moral sensitivity did increase, ultimately making the scientists’ more ethically aware of the actions in their work. 

Our fellowship was aimed at discussing how scientific decision-making includes both epistemic and non-epistemic values, and entanglements between them. We know theoretically that scientific decision-making is value-laden and there are calls to increase scientists' understanding and appreciation of this

One way to improve scientists’ moral behavior is to improve the different moral capacities in the Four Component Model. Since the four component model suggests that people form their ethical decisions based on considering moral sensitivity, moral reasoning, moral motivation, and moral implementation, improving these steps will ultimately lead to scientists being more ethically attentive in their studies. 

In the end we want scientists to be more attentive towards ethical issues throughout their work and to ultimately make better ethical judgements about both research conduct and the broader consequences of their work. Our findings suggest that they actually were more attentive to ethical values after the fellowship. This suggests there is promise in approaches to ethics training that focus on the general role of values in scientific practice.

\begin{acknowledgments}
We would like to thank the Kansas State University Physics Department for their support. We would also like to thank K-SUPER for their valuable insights in our study. This material is based upon work supported by the National Science Foundation under Grant No. \#1835366.
\end{acknowledgments}


\bibliography{GVS}

\begin{thebibliography}{49}%
\makeatletter
\providecommand \@ifxundefined [1]{%
 \@ifx{#1\undefined}
}%
\providecommand \@ifnum [1]{%
 \ifnum #1\expandafter \@firstoftwo
 \else \expandafter \@secondoftwo
 \fi
}%
\providecommand \@ifx [1]{%
 \ifx #1\expandafter \@firstoftwo
 \else \expandafter \@secondoftwo
 \fi
}%
\providecommand \natexlab [1]{#1}%
\providecommand \enquote  [1]{``#1''}%
\providecommand \bibnamefont  [1]{#1}%
\providecommand \bibfnamefont [1]{#1}%
\providecommand \citenamefont [1]{#1}%
\providecommand \href@noop [0]{\@secondoftwo}%
\providecommand \href [0]{\begingroup \@sanitize@url \@href}%
\providecommand \@href[1]{\@@startlink{#1}\@@href}%
\providecommand \@@href[1]{\endgroup#1\@@endlink}%
\providecommand \@sanitize@url [0]{\catcode `\\12\catcode `\$12\catcode
  `\&12\catcode `\#12\catcode `\^12\catcode `\_12\catcode `\%12\relax}%
\providecommand \@@startlink[1]{}%
\providecommand \@@endlink[0]{}%
\providecommand \url  [0]{\begingroup\@sanitize@url \@url }%
\providecommand \@url [1]{\endgroup\@href {#1}{\urlprefix }}%
\providecommand \urlprefix  [0]{URL }%
\providecommand \Eprint [0]{\href }%
\providecommand \doibase [0]{https://doi.org/}%
\providecommand \selectlanguage [0]{\@gobble}%
\providecommand \bibinfo  [0]{\@secondoftwo}%
\providecommand \bibfield  [0]{\@secondoftwo}%
\providecommand \translation [1]{[#1]}%
\providecommand \BibitemOpen [0]{}%
\providecommand \bibitemStop [0]{}%
\providecommand \bibitemNoStop [0]{.\EOS\space}%
\providecommand \EOS [0]{\spacefactor3000\relax}%
\providecommand \BibitemShut  [1]{\csname bibitem#1\endcsname}%
\let\auto@bib@innerbib\@empty
\bibitem [{Gui()}]{Guidlines}%
  \BibitemOpen
  \href {http://www.aps.org/policy/statements/guidlinesethics.cfm}
  {{\selectlanguage {en}\bibinfo {title} {19.1 {Guidelines} on {Ethics} ({Full}
  {Statement})}}}\BibitemShut {NoStop}%
\bibitem [{fos()}]{fostering}%
  \BibitemOpen
  \href {https://doi.org/10.17226/21896} {{\selectlanguage {en}\emph {\bibinfo
  {title} {Read "{Fostering} {Integrity} in {Research}" at
  {NAP}.edu}}}}\BibitemShut {NoStop}%
\bibitem [{\citenamefont {Reiss}(1999)}]{reiss_teaching_1999}%
  \BibitemOpen
  \bibfield  {author} {\bibinfo {author} {\bibfnamefont {M.~J.}\ \bibnamefont
  {Reiss}},\ }\bibfield  {title} {{\selectlanguage {en}\bibinfo {title}
  {Teaching {Ethics} in {Science}}},\ }\href
  {https://doi.org/10.1080/03057269908560151} {\bibfield  {journal} {\bibinfo
  {journal} {Studies in Science Education}\ }\textbf {\bibinfo {volume} {34}},\
  \bibinfo {pages} {115} (\bibinfo {year} {1999})}\BibitemShut {NoStop}%
\bibitem [{CIT()}]{CITI}%
  \BibitemOpen
  \href
  {https://about.citiprogram.org/en/course/responsible-conduct-of-research-basic/}
  {{\selectlanguage {en}\bibinfo {title} {Responsible {Conduct} of {Research}
  ({RCR}) {Basic} {\textbar} {CITI} {Program}}}}\BibitemShut {NoStop}%
\bibitem [{\citenamefont {Mumford}()}]{mumford_read_nodate}%
  \BibitemOpen
  \bibfield  {author} {\bibinfo {author} {\bibfnamefont {M.}~\bibnamefont
  {Mumford}},\ }\href {https://doi.org/10.17226/21896} {{\selectlanguage
  {en}\emph {\bibinfo {title} {Read "{Fostering} {Integrity} in {Research}" at
  {NAP}.edu}}}}\BibitemShut {NoStop}%
\bibitem [{\citenamefont {Powell}\ \emph {et~al.}(2007)\citenamefont {Powell},
  \citenamefont {Allison},\ and\ \citenamefont
  {Kalichman}}]{powell_effectiveness_2007}%
  \BibitemOpen
  \bibfield  {author} {\bibinfo {author} {\bibfnamefont {S.}~\bibnamefont
  {Powell}}, \bibinfo {author} {\bibfnamefont {M.}~\bibnamefont {Allison}},\
  and\ \bibinfo {author} {\bibfnamefont {M.}~\bibnamefont {Kalichman}},\
  }\bibfield  {title} {\bibinfo {title} {Effectiveness of a {Responsible}
  {Conduct} of {Research} {Course}: {A} {Preliminary} {Study}},\ }\href
  {https://doi.org/10.1007/s11948-007-9012-y} {\bibfield  {journal} {\bibinfo
  {journal} {Science and engineering ethics}\ }\textbf {\bibinfo {volume}
  {13}},\ \bibinfo {pages} {249} (\bibinfo {year} {2007})}\BibitemShut
  {NoStop}%
\bibitem [{\citenamefont {Antes}\ \emph {et~al.}(2009)\citenamefont {Antes},
  \citenamefont {Murphy}, \citenamefont {Waples}, \citenamefont {Mumford},
  \citenamefont {Brown}, \citenamefont {Connelly},\ and\ \citenamefont
  {Devenport}}]{antes_meta-analysis_2009}%
  \BibitemOpen
  \bibfield  {author} {\bibinfo {author} {\bibfnamefont {A.~L.}\ \bibnamefont
  {Antes}}, \bibinfo {author} {\bibfnamefont {S.~T.}\ \bibnamefont {Murphy}},
  \bibinfo {author} {\bibfnamefont {E.~P.}\ \bibnamefont {Waples}}, \bibinfo
  {author} {\bibfnamefont {M.~D.}\ \bibnamefont {Mumford}}, \bibinfo {author}
  {\bibfnamefont {R.~P.}\ \bibnamefont {Brown}}, \bibinfo {author}
  {\bibfnamefont {S.}~\bibnamefont {Connelly}},\ and\ \bibinfo {author}
  {\bibfnamefont {L.~D.}\ \bibnamefont {Devenport}},\ }\bibfield  {title}
  {\bibinfo {title} {A {Meta}-{Analysis} of {Ethics} {Instruction}
  {Effectiveness} in the {Sciences}},\ }\href
  {https://doi.org/10.1080/10508420903035380} {\bibfield  {journal} {\bibinfo
  {journal} {Ethics \& Behavior}\ }\textbf {\bibinfo {volume} {19}},\ \bibinfo
  {pages} {379} (\bibinfo {year} {2009})},\ \bibinfo {note} {publisher:
  Routledge, eprint: https://doi.org/10.1080/10508420903035380}\BibitemShut
  {NoStop}%
\bibitem [{\citenamefont {Kirby}\ and\ \citenamefont
  {Houle}(2004)}]{kirby_ethics_2004}%
  \BibitemOpen
  \bibfield  {author} {\bibinfo {author} {\bibfnamefont {K.}~\bibnamefont
  {Kirby}}\ and\ \bibinfo {author} {\bibfnamefont {F.~A.}\ \bibnamefont
  {Houle}},\ }\bibfield  {title} {\bibinfo {title} {Ethics and the {Welfare} of
  the {Physics} {Profession}},\ }\href {https://doi.org/10.1063/1.1839376}
  {\bibfield  {journal} {\bibinfo  {journal} {Physics Today}\ }\textbf
  {\bibinfo {volume} {57}},\ \bibinfo {pages} {42} (\bibinfo {year} {2004})},\
  \bibinfo {note} {publisher: American Institute of Physics}\BibitemShut
  {NoStop}%
\bibitem [{\citenamefont {Houle}\ \emph {et~al.}(2023)\citenamefont {Houle},
  \citenamefont {Kirby},\ and\ \citenamefont {Marder}}]{houle_ethics_2023}%
  \BibitemOpen
  \bibfield  {author} {\bibinfo {author} {\bibfnamefont {F.~A.}\ \bibnamefont
  {Houle}}, \bibinfo {author} {\bibfnamefont {K.~P.}\ \bibnamefont {Kirby}},\
  and\ \bibinfo {author} {\bibfnamefont {M.~P.}\ \bibnamefont {Marder}},\
  }\bibfield  {title} {\bibinfo {title} {Ethics in physics: {The} need for
  culture change},\ }\href {https://doi.org/10.1063/PT.3.5156} {\bibfield
  {journal} {\bibinfo  {journal} {Physics Today}\ }\textbf {\bibinfo {volume}
  {76}},\ \bibinfo {pages} {28} (\bibinfo {year} {2023})},\ \bibinfo {note}
  {publisher: American Institute of Physics}\BibitemShut {NoStop}%
\bibitem [{\citenamefont {Brown}(2013)}]{brown_values_2013}%
  \BibitemOpen
  \bibfield  {author} {\bibinfo {author} {\bibfnamefont {M.~J.}\ \bibnamefont
  {Brown}},\ }\bibfield  {title} {\bibinfo {title} {Values in {Science} beyond
  {Underdetermination} and {Inductive} {Risk}},\ }\href
  {https://doi.org/10.1086/673720} {\bibfield  {journal} {\bibinfo  {journal}
  {Philosophy of Science}\ }\textbf {\bibinfo {volume} {80}},\ \bibinfo {pages}
  {829} (\bibinfo {year} {2013})}\BibitemShut {NoStop}%
\bibitem [{\citenamefont {Douglas}(2000)}]{douglas_inductive_2000}%
  \BibitemOpen
  \bibfield  {author} {\bibinfo {author} {\bibfnamefont {H.}~\bibnamefont
  {Douglas}},\ }\bibfield  {title} {\bibinfo {title} {Inductive {Risk} and
  {Values} in {Science}},\ }\href {https://www.jstor.org/stable/188707}
  {\bibfield  {journal} {\bibinfo  {journal} {Philosophy of Science}\ }\textbf
  {\bibinfo {volume} {67}},\ \bibinfo {pages} {559} (\bibinfo {year} {2000})},\
  \bibinfo {note} {publisher: [The University of Chicago Press, Philosophy of
  Science Association]}\BibitemShut {NoStop}%
\bibitem [{\citenamefont {Stokhof}(2018)}]{stokhof_ethics_2018}%
  \BibitemOpen
  \bibfield  {author} {\bibinfo {author} {\bibfnamefont {M.~J.~B.}\
  \bibnamefont {Stokhof}},\ }\bibfield  {title} {{\selectlanguage {en}\bibinfo
  {title} {Ethics and morality, principles and practice}},\ }\href
  {https://doi.org/10.1007/s42048-018-0016-x} {\bibfield  {journal} {\bibinfo
  {journal} {Zeitschrift für Ethik und Moralphilosophie}\ }\textbf {\bibinfo
  {volume} {1}},\ \bibinfo {pages} {291} (\bibinfo {year} {2018})}\BibitemShut
  {NoStop}%
\bibitem [{\citenamefont {Rest}\ and\ \citenamefont
  {Narváez}(1994)}]{rest_moral_1994}%
  \BibitemOpen
  \bibinfo {editor} {\bibfnamefont {J.~R.}\ \bibnamefont {Rest}}\ and\ \bibinfo
  {editor} {\bibfnamefont {D.}~\bibnamefont {Narváez}},\ eds.,\ \href@noop {}
  {\emph {\bibinfo {title} {Moral development in the professions: {Psychology}
  and applied ethics}}},\ Moral development in the professions: {Psychology}
  and applied ethics\ (\bibinfo  {publisher} {Lawrence Erlbaum Associates,
  Inc},\ \bibinfo {address} {Hillsdale, NJ, US},\ \bibinfo {year} {1994})\
  \bibinfo {note} {pages: xii, 233}\BibitemShut {NoStop}%
\bibitem [{\citenamefont {Narvaez}\ and\ \citenamefont
  {Rest}(1995)}]{narvaez_four_1995}%
  \BibitemOpen
  \bibfield  {author} {\bibinfo {author} {\bibfnamefont {D.}~\bibnamefont
  {Narvaez}}\ and\ \bibinfo {author} {\bibfnamefont {J.}~\bibnamefont {Rest}},\
  }\bibfield  {title} {\bibinfo {title} {The four components of acting morally.
  {Moral} behavior and moral development: {An} introduction},\ }\href@noop {}
  {\bibfield  {journal} {\bibinfo  {journal} {Handbook of moral and character
  education}\ ,\ \bibinfo {pages} {385}} (\bibinfo {year} {1995})}\BibitemShut
  {NoStop}%
\bibitem [{\citenamefont {You}\ and\ \citenamefont
  {Bebeau}(2013)}]{you_independence_2013}%
  \BibitemOpen
  \bibfield  {author} {\bibinfo {author} {\bibfnamefont {D.}~\bibnamefont
  {You}}\ and\ \bibinfo {author} {\bibfnamefont {M.~J.}\ \bibnamefont
  {Bebeau}},\ }\bibfield  {title} {{\selectlanguage {en}\bibinfo {title} {The
  independence of {James} {Rest}'s components of morality: evidence from a
  professional ethics curriculum study}},\ }\href
  {https://doi.org/10.1080/17449642.2013.846059} {\bibfield  {journal}
  {\bibinfo  {journal} {Ethics and Education}\ }\textbf {\bibinfo {volume}
  {8}},\ \bibinfo {pages} {202} (\bibinfo {year} {2013})}\BibitemShut {NoStop}%
\bibitem [{\citenamefont {Elliott}(2022)}]{elliott_values_2022}%
  \BibitemOpen
  \bibfield  {author} {\bibinfo {author} {\bibfnamefont {K.~C.}\ \bibnamefont
  {Elliott}},\ }\bibfield  {title} {{\selectlanguage {en}\bibinfo {title}
  {Values in {Science}}},\ }\bibfield  {journal} {\bibinfo  {journal} {Elements
  in the Philosophy of Science}\ }\href {https://doi.org/10.1017/9781009052597}
  {10.1017/9781009052597} (\bibinfo {year} {2022}),\ \bibinfo {note} {iSBN:
  9781009052597 9781009055635 Publisher: Cambridge University
  Press}\BibitemShut {NoStop}%
\bibitem [{\citenamefont {Bright}(2018)}]{bright_du_2018}%
  \BibitemOpen
  \bibfield  {author} {\bibinfo {author} {\bibfnamefont {L.~K.}\ \bibnamefont
  {Bright}},\ }\bibfield  {title} {{\selectlanguage {en}\bibinfo {title} {Du
  {Bois}’ democratic defence of the value free ideal}},\ }\href
  {https://doi.org/10.1007/s11229-017-1333-z} {\bibfield  {journal} {\bibinfo
  {journal} {Synthese}\ }\textbf {\bibinfo {volume} {195}},\ \bibinfo {pages}
  {2227} (\bibinfo {year} {2018})}\BibitemShut {NoStop}%
\bibitem [{\citenamefont {Rest}\ \emph
  {et~al.}(1999{\natexlab{a}})\citenamefont {Rest}, \citenamefont {Narvaez},
  \citenamefont {Thoma},\ and\ \citenamefont {Bebeau}}]{rest_dit2_1999}%
  \BibitemOpen
  \bibfield  {author} {\bibinfo {author} {\bibfnamefont {J.}~\bibnamefont
  {Rest}}, \bibinfo {author} {\bibfnamefont {D.}~\bibnamefont {Narvaez}},
  \bibinfo {author} {\bibfnamefont {S.}~\bibnamefont {Thoma}},\ and\ \bibinfo
  {author} {\bibfnamefont {M.}~\bibnamefont {Bebeau}},\ }\bibfield  {title}
  {\bibinfo {title} {{DIT2}: {Devising} and {Testing} {A} {Revised}
  {Instrument} of {Moral} {Judgment}},\ }\href
  {https://doi.org/10.1037/0022-0663.91.4.644} {\bibfield  {journal} {\bibinfo
  {journal} {Journal of Educational Psychology}\ }\textbf {\bibinfo {volume}
  {91}},\ \bibinfo {pages} {644} (\bibinfo {year}
  {1999}{\natexlab{a}})}\BibitemShut {NoStop}%
\bibitem [{\citenamefont {Linville}\ \emph {et~al.}(2023)\citenamefont
  {Linville}, \citenamefont {Cairns}, \citenamefont {Garcia}, \citenamefont
  {Bridges}, \citenamefont {Herington}, \citenamefont {Laverty},\ and\
  \citenamefont {Tanona}}]{linville_how_2023}%
  \BibitemOpen
  \bibfield  {author} {\bibinfo {author} {\bibfnamefont {C.~L.}\ \bibnamefont
  {Linville}}, \bibinfo {author} {\bibfnamefont {A.~C.}\ \bibnamefont
  {Cairns}}, \bibinfo {author} {\bibfnamefont {T.}~\bibnamefont {Garcia}},
  \bibinfo {author} {\bibfnamefont {B.}~\bibnamefont {Bridges}}, \bibinfo
  {author} {\bibfnamefont {J.}~\bibnamefont {Herington}}, \bibinfo {author}
  {\bibfnamefont {J.~T.}\ \bibnamefont {Laverty}},\ and\ \bibinfo {author}
  {\bibfnamefont {S.}~\bibnamefont {Tanona}},\ }\bibfield  {title}
  {{\selectlanguage {en}\bibinfo {title} {How {Do} {Scientists} {Perceive} the
  {Relationship} {Between} {Ethics} and {Science}? {A} {Pilot} {Study} of
  {Scientists}’ {Appeals} to {Values}}},\ }\href
  {https://doi.org/10.1007/s11948-023-00429-1} {\bibfield  {journal} {\bibinfo
  {journal} {Science and Engineering Ethics}\ }\textbf {\bibinfo {volume}
  {29}},\ \bibinfo {pages} {15} (\bibinfo {year} {2023})}\BibitemShut {NoStop}%
\bibitem [{\citenamefont {Bebeau}\ \emph {et~al.}(1985)\citenamefont {Bebeau},
  \citenamefont {Rest},\ and\ \citenamefont {Yamoor}}]{bebeau_measuring_1985}%
  \BibitemOpen
  \bibfield  {author} {\bibinfo {author} {\bibfnamefont {M.}~\bibnamefont
  {Bebeau}}, \bibinfo {author} {\bibfnamefont {J.}~\bibnamefont {Rest}},\ and\
  \bibinfo {author} {\bibfnamefont {C.}~\bibnamefont {Yamoor}},\ }\bibfield
  {title} {\bibinfo {title} {Measuring dental students' ethical sensitivity},\
  }\href {https://doi.org/10.1002/j.0022-0337.1985.49.4.tb01874.x} {\bibfield
  {journal} {\bibinfo  {journal} {Journal of dental education}\ }\textbf
  {\bibinfo {volume} {49}},\ \bibinfo {pages} {225} (\bibinfo {year}
  {1985})}\BibitemShut {NoStop}%
\bibitem [{\citenamefont {Hébert}\ \emph {et~al.}(1992)\citenamefont
  {Hébert}, \citenamefont {Meslin},\ and\ \citenamefont
  {Dunn}}]{hebert_measuring_1992}%
  \BibitemOpen
  \bibfield  {author} {\bibinfo {author} {\bibfnamefont {P.~C.}\ \bibnamefont
  {Hébert}}, \bibinfo {author} {\bibfnamefont {E.~M.}\ \bibnamefont
  {Meslin}},\ and\ \bibinfo {author} {\bibfnamefont {E.~V.}\ \bibnamefont
  {Dunn}},\ }\bibfield  {title} {{\selectlanguage {en}\bibinfo {title}
  {Measuring the ethical sensitivity of medical students: a study at the
  {University} of {Toronto}}},\ }\href {https://doi.org/10.1136/jme.18.3.142}
  {\bibfield  {journal} {\bibinfo  {journal} {Journal of Medical Ethics}\
  }\textbf {\bibinfo {volume} {18}},\ \bibinfo {pages} {142} (\bibinfo {year}
  {1992})}\BibitemShut {NoStop}%
\bibitem [{Abo()}]{About_the_DIT}%
  \BibitemOpen
  \href {https://ethicaldevelopment.ua.edu/about-the-dit.html}
  {{\selectlanguage {en}\bibinfo {title} {About the {DIT}}}}\BibitemShut
  {NoStop}%
\bibitem [{\citenamefont {Rest}\ \emph
  {et~al.}(1999{\natexlab{b}})\citenamefont {Rest}, \citenamefont {Narvaez},
  \citenamefont {Bebeau},\ and\ \citenamefont
  {Thoma}}]{rest_neo-kohlbergian_1999}%
  \BibitemOpen
  \bibfield  {author} {\bibinfo {author} {\bibfnamefont {J.}~\bibnamefont
  {Rest}}, \bibinfo {author} {\bibfnamefont {D.}~\bibnamefont {Narvaez}},
  \bibinfo {author} {\bibfnamefont {M.}~\bibnamefont {Bebeau}},\ and\ \bibinfo
  {author} {\bibfnamefont {S.}~\bibnamefont {Thoma}},\ }\bibfield  {title}
  {\bibinfo {title} {A neo-{Kohlbergian} approach: {The} {DIT} and schema
  theory},\ }\href {https://doi.org/10.1023/A:1022053215271} {\bibfield
  {journal} {\bibinfo  {journal} {Educational Psychology Review}\ }\textbf
  {\bibinfo {volume} {11}},\ \bibinfo {pages} {291} (\bibinfo {year}
  {1999}{\natexlab{b}})},\ \bibinfo {note} {place: Germany Publisher:
  Springer}\BibitemShut {NoStop}%
\bibitem [{\citenamefont {Mumford}\ \emph {et~al.}(2006)\citenamefont
  {Mumford}, \citenamefont {Devenport}, \citenamefont {Brown}, \citenamefont
  {Connelly}, \citenamefont {Murphy}, \citenamefont {Hill},\ and\ \citenamefont
  {Antes}}]{mumford_articles_2006}%
  \BibitemOpen
  \bibfield  {author} {\bibinfo {author} {\bibfnamefont {M.~D.}\ \bibnamefont
  {Mumford}}, \bibinfo {author} {\bibfnamefont {L.~D.}\ \bibnamefont
  {Devenport}}, \bibinfo {author} {\bibfnamefont {R.~P.}\ \bibnamefont
  {Brown}}, \bibinfo {author} {\bibfnamefont {S.}~\bibnamefont {Connelly}},
  \bibinfo {author} {\bibfnamefont {S.~T.}\ \bibnamefont {Murphy}}, \bibinfo
  {author} {\bibfnamefont {J.~H.}\ \bibnamefont {Hill}},\ and\ \bibinfo
  {author} {\bibfnamefont {A.~L.}\ \bibnamefont {Antes}},\ }\bibfield  {title}
  {\bibinfo {title} {{ARTICLES}: {Validation} of {Ethical} {Decision} {Making}
  {Measures}: {Evidence} for a {New} {Set} of {Measures}},\ }\href
  {https://doi.org/10.1207/s15327019eb1604_4} {\bibfield  {journal} {\bibinfo
  {journal} {Ethics \& Behavior}\ }\textbf {\bibinfo {volume} {16}},\ \bibinfo
  {pages} {319} (\bibinfo {year} {2006})},\ \bibinfo {note} {publisher:
  Routledge \_eprint: https://doi.org/10.1207/s15327019eb1604\_4}\BibitemShut
  {NoStop}%
\bibitem [{\citenamefont {Klinker}\ and\ \citenamefont
  {Hackmann}(2004)}]{klinker_analysis_2004}%
  \BibitemOpen
  \bibfield  {author} {\bibinfo {author} {\bibfnamefont {J.~F.}\ \bibnamefont
  {Klinker}}\ and\ \bibinfo {author} {\bibfnamefont {D.~G.}\ \bibnamefont
  {Hackmann}},\ }\bibfield  {title} {{\selectlanguage {en}\bibinfo {title} {An
  {Analysis} of {Principals}’ {Ethical} {Decision} {Making} {Using} {Rest}'s
  {Four} {Component} {Model} of {Moral} {Behavior}}},\ }\href
  {https://doi.org/10.1177/105268460401400404} {\bibfield  {journal} {\bibinfo
  {journal} {Journal of School Leadership}\ }\textbf {\bibinfo {volume} {14}},\
  \bibinfo {pages} {434} (\bibinfo {year} {2004})},\ \bibinfo {note}
  {publisher: SAGE Publications Inc}\BibitemShut {NoStop}%
\bibitem [{\citenamefont {Bebeau}\ \emph {et~al.}(1993)\citenamefont {Bebeau},
  \citenamefont {Born},\ and\ \citenamefont {Ozar}}]{bebeau_development_1993}%
  \BibitemOpen
  \bibfield  {author} {\bibinfo {author} {\bibfnamefont {M.~J.}\ \bibnamefont
  {Bebeau}}, \bibinfo {author} {\bibfnamefont {D.~O.}\ \bibnamefont {Born}},\
  and\ \bibinfo {author} {\bibfnamefont {D.~T.}\ \bibnamefont {Ozar}},\
  }\bibfield  {title} {{\selectlanguage {en}\bibinfo {title} {The development
  of a professional role orientation inventory}},\ }\href@noop {} {\bibfield
  {journal} {\bibinfo  {journal} {The Journal of the American College of
  Dentists}\ }\textbf {\bibinfo {volume} {60}},\ \bibinfo {pages} {27}
  (\bibinfo {year} {1993})}\BibitemShut {NoStop}%
\bibitem [{\citenamefont {Phillips}\ \emph {et~al.}(2018)\citenamefont
  {Phillips}, \citenamefont {Nestor}, \citenamefont {Beach},\ and\
  \citenamefont {Heitman}}]{phillips_america_2018}%
  \BibitemOpen
  \bibfield  {author} {\bibinfo {author} {\bibfnamefont {T.}~\bibnamefont
  {Phillips}}, \bibinfo {author} {\bibfnamefont {F.}~\bibnamefont {Nestor}},
  \bibinfo {author} {\bibfnamefont {G.}~\bibnamefont {Beach}},\ and\ \bibinfo
  {author} {\bibfnamefont {E.}~\bibnamefont {Heitman}},\ }\bibfield  {title}
  {\bibinfo {title} {America {Competes} at 5 {Years}: {An} {Analysis} of
  {Research}-{Intensive} {Universities}? {Rcr} {Training} {Plans}},\ }\href
  {https://doi.org/10.1007/s11948-017-9883-5} {\bibfield  {journal} {\bibinfo
  {journal} {Science and Engineering Ethics}\ }\textbf {\bibinfo {volume}
  {24}},\ \bibinfo {pages} {227} (\bibinfo {year} {2018})},\ \bibinfo {note}
  {publisher: Springer Verlag}\BibitemShut {NoStop}%
\bibitem [{\citenamefont {Dubois}\ and\ \citenamefont
  {Dueker}(2009)}]{dubois_teaching_2009}%
  \BibitemOpen
  \bibfield  {author} {\bibinfo {author} {\bibfnamefont {J.~M.}\ \bibnamefont
  {Dubois}}\ and\ \bibinfo {author} {\bibfnamefont {J.~M.}\ \bibnamefont
  {Dueker}},\ }\bibfield  {title} {{\selectlanguage {en}\bibinfo {title}
  {Teaching and {Assessing} the {Responsible} {Conduct} of {Research}: {A}
  {Delphi} {Consensus} {Panel} {Report}}},\ }\href@noop {} {\bibfield
  {journal} {\bibinfo  {journal} {The Journal of Research Administration}\
  }\textbf {\bibinfo {volume} {40}},\ \bibinfo {pages} {49} (\bibinfo {year}
  {2009})}\BibitemShut {NoStop}%
\bibitem [{\citenamefont {Kalichman}(2007)}]{kalichman_responding_2007}%
  \BibitemOpen
  \bibfield  {author} {\bibinfo {author} {\bibfnamefont {M.~W.}\ \bibnamefont
  {Kalichman}},\ }\bibfield  {title} {{\selectlanguage {en}\bibinfo {title}
  {Responding to {Challenges} in {Educating} for the {Responsible} {Conduct} of
  {Research}}},\ }\href {https://doi.org/10.1097/ACM.0b013e31812f77fe}
  {\bibfield  {journal} {\bibinfo  {journal} {Academic Medicine}\ }\textbf
  {\bibinfo {volume} {82}},\ \bibinfo {pages} {870} (\bibinfo {year}
  {2007})}\BibitemShut {NoStop}%
\bibitem [{\citenamefont {Pennock}\ and\ \citenamefont
  {O'Rourke}(2017)}]{pennock_developing_2017}%
  \BibitemOpen
  \bibfield  {author} {\bibinfo {author} {\bibfnamefont {R.}~\bibnamefont
  {Pennock}}\ and\ \bibinfo {author} {\bibfnamefont {M.}~\bibnamefont
  {O'Rourke}},\ }\bibfield  {title} {\bibinfo {title} {Developing a
  {Scientific} {Virtue}-{Based} {Approach} to {Science} {Ethics} {Training}},\
  }\bibfield  {journal} {\bibinfo  {journal} {Science and engineering ethics}\
  }\textbf {\bibinfo {volume} {23}},\ \href
  {https://doi.org/10.1007/s11948-016-9757-2} {10.1007/s11948-016-9757-2}
  (\bibinfo {year} {2017})\BibitemShut {NoStop}%
\bibitem [{\citenamefont {Cairns}\ \emph {et~al.}(2021)\citenamefont {Cairns},
  \citenamefont {Linville}, \citenamefont {Garcia}, \citenamefont {Bridges},
  \citenamefont {Tanona}, \citenamefont {Herington},\ and\ \citenamefont
  {Laverty}}]{cairns_phenomenographic_2021}%
  \BibitemOpen
  \bibfield  {author} {\bibinfo {author} {\bibfnamefont {A.~C.}\ \bibnamefont
  {Cairns}}, \bibinfo {author} {\bibfnamefont {C.}~\bibnamefont {Linville}},
  \bibinfo {author} {\bibfnamefont {T.}~\bibnamefont {Garcia}}, \bibinfo
  {author} {\bibfnamefont {B.}~\bibnamefont {Bridges}}, \bibinfo {author}
  {\bibfnamefont {S.}~\bibnamefont {Tanona}}, \bibinfo {author} {\bibfnamefont
  {J.}~\bibnamefont {Herington}},\ and\ \bibinfo {author} {\bibfnamefont
  {J.~T.}\ \bibnamefont {Laverty}},\ }\bibfield  {title} {{\selectlanguage
  {en}\bibinfo {title} {A phenomenographic study of scientists’ beliefs about
  the causes of scientists’ research misconduct}},\ }\href
  {https://doi.org/10.1177/17470161211042658} {\bibfield  {journal} {\bibinfo
  {journal} {Research Ethics}\ ,\ \bibinfo {pages} {17470161211042658}}
  (\bibinfo {year} {2021})},\ \bibinfo {note} {publisher: SAGE Publications
  Ltd}\BibitemShut {NoStop}%
\bibitem [{\citenamefont {Hausman}(2011)}]{hausman_2011}%
  \BibitemOpen
  \bibfield  {author} {\bibinfo {author} {\bibfnamefont {D.~M.}\ \bibnamefont
  {Hausman}},\ }\href {https://doi.org/10.1017/CBO9781139058537} {\emph
  {\bibinfo {title} {Preference, Value, Choice, and Welfare}}}\ (\bibinfo
  {publisher} {Cambridge University Press},\ \bibinfo {year}
  {2011})\BibitemShut {NoStop}%
\bibitem [{\citenamefont {Kagan}(2018)}]{kagan_normative_2018}%
  \BibitemOpen
  \bibfield  {author} {\bibinfo {author} {\bibfnamefont {S.}~\bibnamefont
  {Kagan}},\ }\href@noop {} {{\selectlanguage {en}\emph {\bibinfo {title}
  {Normative {Ethics}}}}}\ (\bibinfo  {publisher} {Routledge},\ \bibinfo {year}
  {2018})\ \bibinfo {note} {google-Books-ID: 7f\_EDwAAQBAJ}\BibitemShut
  {NoStop}%
\bibitem [{\citenamefont {Rest}(1984)}]{rest_research_1984}%
  \BibitemOpen
  \bibfield  {author} {\bibinfo {author} {\bibfnamefont {J.~R.}\ \bibnamefont
  {Rest}},\ }\bibfield  {title} {{\selectlanguage {en}\bibinfo {title}
  {Research on {Moral} {Development}: {Implications} for {Training}
  {Counseling} {Psychologists}}},\ }\href
  {https://doi.org/10.1177/0011000084123003} {\bibfield  {journal} {\bibinfo
  {journal} {The Counseling Psychologist}\ }\textbf {\bibinfo {volume} {12}},\
  \bibinfo {pages} {19} (\bibinfo {year} {1984})},\ \bibinfo {note} {publisher:
  SAGE Publications Inc}\BibitemShut {NoStop}%
\bibitem [{\citenamefont {Thoma}(2014{\natexlab{a}})}]{thoma_defining_nodate}%
  \BibitemOpen
  \bibfield  {author} {\bibinfo {author} {\bibfnamefont {S.~J.}\ \bibnamefont
  {Thoma}},\ }\bibfield  {title} {{\selectlanguage {en}\bibinfo {title} {The
  {Defining} {Issues} {Test} of moral judgment development.}},\ }\href
  {https://doi.org/10.1037/h0100590} {\bibfield  {journal} {\bibinfo  {journal}
  {Behavioral Development Bulletin}\ }\textbf {\bibinfo {volume} {19}},\
  \bibinfo {pages} {55} (\bibinfo {year} {2014}{\natexlab{a}})},\ \bibinfo
  {note} {publisher: US: Joseph D. Cautilli}\BibitemShut {NoStop}%
\bibitem [{\citenamefont {Rest}\ \emph
  {et~al.}(1999{\natexlab{c}})\citenamefont {Rest}, \citenamefont {Narvaez},
  \citenamefont {Bebeau},\ and\ \citenamefont
  {Thoma}}]{rest_postconventional_1999}%
  \BibitemOpen
  \bibfield  {author} {\bibinfo {author} {\bibfnamefont {J.}~\bibnamefont
  {Rest}}, \bibinfo {author} {\bibfnamefont {D.}~\bibnamefont {Narvaez}},
  \bibinfo {author} {\bibfnamefont {M.~J.}\ \bibnamefont {Bebeau}},\ and\
  \bibinfo {author} {\bibfnamefont {S.~J.}\ \bibnamefont {Thoma}},\ }\href@noop
  {} {\emph {\bibinfo {title} {Postconventional moral thinking: {A}
  neo-{Kohlbergian} approach}}},\ Postconventional moral thinking: {A}
  neo-{Kohlbergian} approach\ (\bibinfo  {publisher} {Lawrence Erlbaum
  Associates Publishers},\ \bibinfo {address} {Mahwah, NJ, US},\ \bibinfo
  {year} {1999})\ \bibinfo {note} {pages: ix, 229}\BibitemShut {NoStop}%
\bibitem [{\citenamefont {Thoma}(2014{\natexlab{b}})}]{thoma_measuring_2014}%
  \BibitemOpen
  \bibfield  {author} {\bibinfo {author} {\bibfnamefont {S.~J.}\ \bibnamefont
  {Thoma}},\ }\bibfield  {title} {{\selectlanguage {en}\bibinfo {title}
  {Measuring moral thinking from a neo-{Kohlbergian} perspective}},\ }\href
  {https://doi.org/10.1177/1477878514545208} {\bibfield  {journal} {\bibinfo
  {journal} {Theory and Research in Education}\ }\textbf {\bibinfo {volume}
  {12}},\ \bibinfo {pages} {347} (\bibinfo {year}
  {2014}{\natexlab{b}})}\BibitemShut {NoStop}%
\bibitem [{\citenamefont {Myyry}\ and\ \citenamefont
  {Helkama}(2002)}]{myyry_role_2002}%
  \BibitemOpen
  \bibfield  {author} {\bibinfo {author} {\bibfnamefont {L.}~\bibnamefont
  {Myyry}}\ and\ \bibinfo {author} {\bibfnamefont {K.}~\bibnamefont
  {Helkama}},\ }\bibfield  {title} {\bibinfo {title} {The {Role} of {Value}
  {Priorities} and {Professional} {Ethics} {Training} in {Moral}
  {Sensitivity}},\ }\href {https://doi.org/10.1080/03057240120111427}
  {\bibfield  {journal} {\bibinfo  {journal} {Journal of Moral Education}\
  }\textbf {\bibinfo {volume} {31}},\ \bibinfo {pages} {35} (\bibinfo {year}
  {2002})},\ \bibinfo {note} {publisher: Routledge \_eprint:
  https://doi.org/10.1080/03057240120111427}\BibitemShut {NoStop}%
\bibitem [{\citenamefont {Clarkeburn}\ \emph {et~al.}(2002)\citenamefont
  {Clarkeburn}, \citenamefont {Downie},\ and\ \citenamefont
  {Matthew}}]{clarkeburn_impact_2002}%
  \BibitemOpen
  \bibfield  {author} {\bibinfo {author} {\bibfnamefont {H.}~\bibnamefont
  {Clarkeburn}}, \bibinfo {author} {\bibfnamefont {J.~R.}\ \bibnamefont
  {Downie}},\ and\ \bibinfo {author} {\bibfnamefont {B.}~\bibnamefont
  {Matthew}},\ }\bibfield  {title} {\bibinfo {title} {Impact of an {Ethics}
  {Programme} in a {Life} {Sciences} {Curriculum}},\ }\href
  {https://doi.org/10.1080/13562510120100391} {\bibfield  {journal} {\bibinfo
  {journal} {Teaching in Higher Education}\ }\textbf {\bibinfo {volume} {7}},\
  \bibinfo {pages} {65} (\bibinfo {year} {2002})},\ \bibinfo {note} {publisher:
  Routledge \_eprint: https://doi.org/10.1080/13562510120100391}\BibitemShut
  {NoStop}%
\bibitem [{\citenamefont {Kohlberg}\ \emph {et~al.}(1983)\citenamefont
  {Kohlberg}, \citenamefont {Levine}, ,\ and\ \citenamefont
  {Hewer}}]{kohlberg_moral_1983}%
  \BibitemOpen
  \bibfield  {author} {\bibinfo {author} {\bibfnamefont {L.}~\bibnamefont
  {Kohlberg}}, \bibinfo {author} {\bibfnamefont {C.}~\bibnamefont {Levine}}, ,\
  and\ \bibinfo {author} {\bibfnamefont {A.}~\bibnamefont {Hewer}},\ }\href
  {https://repository.library.georgetown.edu/handle/10822/796521}
  {{\selectlanguage {en}\bibinfo {title} {Moral {Stages}: {A} {Current}
  {Formulation} and a {Response} to {Critics}}}} (\bibinfo {year} {1983}),\
  \bibinfo {note} {accepted: 2016-01-08T18:51:08Z}\BibitemShut {NoStop}%
\bibitem [{\citenamefont {Hsu}\ and\ \citenamefont
  {Lachenbruch}(2008)}]{hsu_paired_2008}%
  \BibitemOpen
  \bibfield  {author} {\bibinfo {author} {\bibfnamefont {H.}~\bibnamefont
  {Hsu}}\ and\ \bibinfo {author} {\bibfnamefont {P.}~\bibnamefont
  {Lachenbruch}},\ }\bibfield  {title} {\bibinfo {title} {Paired t {Test}}\
  }(\bibinfo {year} {2008})\BibitemShut {NoStop}%
\bibitem [{\citenamefont {Lakens}(2013)}]{lakens_calculating_2013}%
  \BibitemOpen
  \bibfield  {author} {\bibinfo {author} {\bibfnamefont {D.}~\bibnamefont
  {Lakens}},\ }\bibfield  {title} {\bibinfo {title} {Calculating and reporting
  effect sizes to facilitate cumulative science: a practical primer for t-tests
  and {ANOVAs}},\ }\href
  {https://www.frontiersin.org/articles/10.3389/fpsyg.2013.00863} {\bibfield
  {journal} {\bibinfo  {journal} {Frontiers in Psychology}\ }\textbf {\bibinfo
  {volume} {4}} (\bibinfo {year} {2013})}\BibitemShut {NoStop}%
\bibitem [{\citenamefont {Bewick}\ \emph {et~al.}(2004)\citenamefont {Bewick},
  \citenamefont {Cheek},\ and\ \citenamefont {Ball}}]{bewick_statistics_2004}%
  \BibitemOpen
  \bibfield  {author} {\bibinfo {author} {\bibfnamefont {V.}~\bibnamefont
  {Bewick}}, \bibinfo {author} {\bibfnamefont {L.}~\bibnamefont {Cheek}},\ and\
  \bibinfo {author} {\bibfnamefont {J.}~\bibnamefont {Ball}},\ }\bibfield
  {title} {\bibinfo {title} {Statistics review 8: {Qualitative} data – tests
  of association},\ }\href {https://doi.org/10.1186/cc2428} {\bibfield
  {journal} {\bibinfo  {journal} {Critical Care}\ }\textbf {\bibinfo {volume}
  {8}},\ \bibinfo {pages} {46} (\bibinfo {year} {2004})}\BibitemShut {NoStop}%
\bibitem [{\citenamefont {Kim}(2017)}]{kim_statistical_2017}%
  \BibitemOpen
  \bibfield  {author} {\bibinfo {author} {\bibfnamefont {H.-Y.}\ \bibnamefont
  {Kim}},\ }\bibfield  {title} {\bibinfo {title} {Statistical notes for
  clinical researchers: {Chi}-squared test and {Fisher}'s exact test},\ }\href
  {https://doi.org/10.5395/rde.2017.42.2.152} {\bibfield  {journal} {\bibinfo
  {journal} {Restorative Dentistry \& Endodontics}\ }\textbf {\bibinfo {volume}
  {42}},\ \bibinfo {pages} {152} (\bibinfo {year} {2017})},\ \bibinfo {note}
  {publisher: The Korean Academy of Conservative Dentistry}\BibitemShut
  {NoStop}%
\bibitem [{\citenamefont {Fleiss}(1971)}]{fleiss_measuring_1971}%
  \BibitemOpen
  \bibfield  {author} {\bibinfo {author} {\bibfnamefont {J.~L.}\ \bibnamefont
  {Fleiss}},\ }\bibfield  {title} {\bibinfo {title} {Measuring nominal scale
  agreement among many raters},\ }\href {https://doi.org/10.1037/h0031619}
  {\bibfield  {journal} {\bibinfo  {journal} {Psychological Bulletin}\ }\textbf
  {\bibinfo {volume} {76}},\ \bibinfo {pages} {378} (\bibinfo {year} {1971})},\
  \bibinfo {note} {place: US Publisher: American Psychological
  Association}\BibitemShut {NoStop}%
\bibitem [{\citenamefont {Hallgren}(2012)}]{hallgren_computing_2012}%
  \BibitemOpen
  \bibfield  {author} {\bibinfo {author} {\bibfnamefont {K.~A.}\ \bibnamefont
  {Hallgren}},\ }\bibfield  {title} {\bibinfo {title} {Computing
  {Inter}-{Rater} {Reliability} for {Observational} {Data}: {An} {Overview} and
  {Tutorial}},\ }\href {https://www.ncbi.nlm.nih.gov/pmc/articles/PMC3402032/}
  {\bibfield  {journal} {\bibinfo  {journal} {Tutorials in quantitative methods
  for psychology}\ }\textbf {\bibinfo {volume} {8}},\ \bibinfo {pages} {23}
  (\bibinfo {year} {2012})}\BibitemShut {NoStop}%
\bibitem [{\citenamefont {Sim}\ and\ \citenamefont
  {Wright}(2005)}]{sim_kappa_2005}%
  \BibitemOpen
  \bibfield  {author} {\bibinfo {author} {\bibfnamefont {J.}~\bibnamefont
  {Sim}}\ and\ \bibinfo {author} {\bibfnamefont {C.~C.}\ \bibnamefont
  {Wright}},\ }\bibfield  {title} {\bibinfo {title} {The {Kappa} {Statistic} in
  {Reliability} {Studies}: {Use}, {Interpretation}, and {Sample} {Size}
  {Requirements}},\ }\href {https://doi.org/10.1093/ptj/85.3.257} {\bibfield
  {journal} {\bibinfo  {journal} {Physical Therapy}\ }\textbf {\bibinfo
  {volume} {85}},\ \bibinfo {pages} {257} (\bibinfo {year} {2005})}\BibitemShut
  {NoStop}%
\bibitem [{\citenamefont {Landis}\ and\ \citenamefont
  {Koch}(1977)}]{landis_measurement_1977}%
  \BibitemOpen
  \bibfield  {author} {\bibinfo {author} {\bibfnamefont {J.~R.}\ \bibnamefont
  {Landis}}\ and\ \bibinfo {author} {\bibfnamefont {G.~G.}\ \bibnamefont
  {Koch}},\ }\bibfield  {title} {{\selectlanguage {en}\bibinfo {title} {The
  {Measurement} of {Observer} {Agreement} for {Categorical} {Data}}},\ }\href
  {https://doi.org/10.2307/2529310} {\bibfield  {journal} {\bibinfo  {journal}
  {Biometrics}\ }\textbf {\bibinfo {volume} {33}},\ \bibinfo {pages} {159}
  (\bibinfo {year} {1977})}\BibitemShut {NoStop}%
\bibitem [{\citenamefont {Garcia}\ \emph {et~al.}(2022)\citenamefont {Garcia},
  \citenamefont {Solis}, \citenamefont {Linville}, \citenamefont {Bridges},
  \citenamefont {Jones}, \citenamefont {Herington}, \citenamefont {Tanona},\
  and\ \citenamefont {Laverty}}]{garcia_examining_2022}%
  \BibitemOpen
  \bibfield  {author} {\bibinfo {author} {\bibfnamefont {T.}~\bibnamefont
  {Garcia}}, \bibinfo {author} {\bibfnamefont {C.}~\bibnamefont {Solis}},
  \bibinfo {author} {\bibfnamefont {C.}~\bibnamefont {Linville}}, \bibinfo
  {author} {\bibfnamefont {B.}~\bibnamefont {Bridges}}, \bibinfo {author}
  {\bibfnamefont {W.}~\bibnamefont {Jones}}, \bibinfo {author} {\bibfnamefont
  {J.}~\bibnamefont {Herington}}, \bibinfo {author} {\bibfnamefont
  {S.}~\bibnamefont {Tanona}},\ and\ \bibinfo {author} {\bibfnamefont {J.~T.}\
  \bibnamefont {Laverty}},\ }\bibfield  {title} {{\selectlanguage {en}\bibinfo
  {title} {Examining {Physicists}' {Ethical} {Reasoning}: {A} {New}
  {Methodology}}}\ }(\bibinfo {year} {2022})\ pp.\ \bibinfo {pages}
  {200--205},\ \bibinfo {note} {iSSN: 2377-2379}\BibitemShut {NoStop}%
\end{thebibliography}%

\end{document}